\documentstyle[psfig,epsfig,11pt]{article}
\input amssym.def
\input amssym.tex
\newtheorem{theorem}{Theorem}
\newtheorem{lemma}{Lemma}

\def\binom#1#2{{#1}\choose{#2}}

\def\slfrac#1#2{\hbox{\kern.1em %
 \raise.5ex\hbox{\the\scriptfont0 #1}\kern-.11em %
 /\kern-.15em\lower.25ex\hbox{\the\scriptfont0 #2}}}

\newcommand{\eqn}[1]{(\ref{#1})}
\newcommand{\hsp}{\hspace*{\parindent}}
\newcommand{\vsp}{\vspace{.1in}}

\newcommand{\eeq}{\end{equation}}
\newcommand{\beql}[1]{\begin{equation}\label{#1}}

\newcommand{\sS}{{\cal S}}

\newcommand{\ZZ}{{\Bbb Z}}
\newcommand{\RR}{{\Bbb R}}

\newcommand{\CC}{{\Bbb C}}

\newcommand{\af}{\alpha}
\newcommand{\la}{\lambda}
\newcommand{\ra}{\rangle}
\newcommand{\om}{\omega}
\newcommand{\oom}{\bar{\omega}}

\newcommand{\ov}{\bar{v}}
\newcommand{\ou}{\bar{u}}

\newcommand{\oS}{\bar{S}}
\newcommand{\oZ}{\bar{Z}}

\newcommand{\os}{\bar{s}}
\newcommand{\oX}{\bar{X}}

\newcommand{\oE}{\bar{E}}

\newcommand{\ds}{\displaystyle\sum}

\newcommand{\sG}{{\cal G}}

\newcommand{\lf}{\lfloor}
\newcommand{\rf}{\rfloor}

\makeatletter
\def\@sect#1#2#3#4#5#6[#7]#8{\ifnum #2>\c@secnumdepth
     \def\@svsec{}\else
     \refstepcounter{#1}\edef\@svsec{\csname the#1\endcsname.\hskip .75em }\fi
     \@tempskipa #5\relax
      \ifdim \@tempskipa>\z@
        \begingroup #6\relax
          \@hangfrom{\hskip #3\relax\@svsec}{\interlinepenalty \@M #8\par}%
        \endgroup
       \csname #1mark\endcsname{#7}\addcontentsline
         {toc}{#1}{\ifnum #2>\c@secnumdepth \else
                      \protect\numberline{\csname the#1\endcsname}\fi
                    #7}\else
        \def\@svsechd{#6\hskip #3\@svsec #8\csname #1mark\endcsname
                      {#7}\addcontentsline
                           {toc}{#1}{\ifnum #2>\c@secnumdepth \else
                             \protect\numberline{\csname the#1\endcsname}\fi
                       #7}}\fi
     \@xsect{#5}}
\def\@begintheorem#1#2{\it \trivlist \item[\hskip \labelsep{\bf #1\ #2.}]}

\def\plain{plain}\ifx\fmtname\plain\csname fi\endcsname
     
     \input docstrip
     \preamble

     Do not distribute the stripped version of this file.
     The checksum in the header refers to the documented version.

     \endpreamble
     \generateFile{here.sty}{t}{\from{here.doc}{}}
     \endinput
\fi
\ifcat a\noexpand @\let\next\relax\else\def\next{%
    \documentstyle[here,doc]{article}\MakePercentIgnore}\fi\next
\ifx\@Hxfloat\@Hundef\else\expandafter\endinput\fi
\let\@Hxfloat\@xfloat
\def\@xfloat#1[{\@ifnextchar{H}{\@HHfloat{#1}[}{\@Hxfloat{#1}[}}
\def\@HHfloat#1[H]{%
\expandafter\let\csname end#1\endcsname\end@Hfloat
\vskip\intextsep\vbox\bgroup\def\@captype{#1}\parindent\z@
\ignorespaces}
\def\end@Hfloat{\egroup\vskip \intextsep}

\makeatother

\begin{document}
\begin{center}
{\Large {\bf Quantum Error Correction Via Codes Over $GF(4)$}} \\
\vspace{1.5\baselineskip}
{\em A. R. Calderbank},$^1$ {\em E. M. Rains},$^2$ {\em P. W. Shor},$^1$ 
and {\em N. J. A. Sloane}$^1$ \\
\vspace*{1\baselineskip}
$^1$AT\&T Labs - Research, Florham Park, New Jersey 07932-0971 \\
\vspace{1\baselineskip}
$^2$Institute for Defense Analyses, Princeton, New Jersey 08540 \\
\vspace{1.5\baselineskip}
August 27, 1997 \\
\vspace{1.5\baselineskip}
{\bf ABSTRACT}
\vspace{.5\baselineskip}
\end{center}

The problem of finding quantum-error-correcting codes is transformed into the problem of
finding additive codes over the field
$GF(4)$ which are self-orthogonal with respect to a certain trace inner product.
Many new codes and new bounds are presented, as well as a table of upper and lower bounds on
such codes of length up to 30 qubits.

\vspace*{+4.5in}
\noindent
{\footnotesize
Manuscript received $\underline{~~~~~~~~~~~~~~~~~~~~~~~~~~~}$; revised $\underline{~~~~~~~~~~~~~~~~~~~~~~~~~~~}$

\vspace*{+.2in}
\noindent
The authors are with AT\&T Labs - Research, Florham Park, NJ 07932-0971, USA.
The work of the second author was performed while he was with the Institute for
Defense Analyses, Princeton, NJ, USA.
}

\clearpage
\thispagestyle{empty}
\setcounter{page}{1}
\setlength{\baselineskip}{1.5\baselineskip}

\section{Introduction}
\hsp
The relationship between quantum information and classical information
is a subject currently receiving much study.
While there are many similarities, there are also substantial differences between
the two.
Classical information cannot travel faster than light,
while quantum information appears to in some circumstances
(although proper definitions can resolve this apparent paradox).
Classical information can be duplicated, while quantum information cannot \cite{Die82}, \cite{WoZu82}.

It is well known that classical information can be protected from
degradation by the use of classical error-correcting codes \cite{MS77}.
Classical error-correcting codes appear to protect classical information by
duplicating it, so because of the theorem that a quantum bit cannot be
cloned, it was widely believed that these techniques could not be applied
to quantum information.  That quantum-error-correcting codes could indeed
exist was recently shown by one of us \cite{Sho95}.  Two of us
\cite{CaSh96} then showed that a class of good quantum codes could be
obtained by using a construction that starts with a binary linear code $C$
containing its dual $C^\perp$.  Independently, Steane also discovered the
existence of quantum codes \cite{Ste96} and the same construction
\cite{SteR}.  At around the same time, Bennett et~al. \cite{BBP96}
discovered that two experimenters each holding one component of many noisy
Einstein-Podolsky-Rosen (EPR) pairs could {\em purify} them using only
a classical channel to obtain fewer nearly perfect EPR pairs.  The
resulting pairs can then be used to teleport quantum information from one
experimenter to the other \cite{BBC93}.  Although it was not immediately
apparent, these two discoveries turned out to be different ways of looking
at the same phenomenon.  A purification protocol that uses only a one-way
classical channel between the experimenters can be converted into a
quantum-error-correcting code, and vice versa \cite{BDSW96}.  After these
discoveries, a number of improved quantum codes were soon found by various
researchers.

The setting in which quantum-error-correcting codes exist is the quantum
state space of $n$ qubits (quantum bits, or two-state quantum systems).
This space is $\CC^{2^n}$, and it has a natural decomposition as the tensor
product of $n$ copies of $\CC^2$, where each copy corresponds to one qubit.
We noticed that the known quantum codes seemed to have close connections to
a finite group of unitary transformations of $\CC^{2^n}$, known as a
Clifford group, and denoted here by $L$.  This group contains all the
transformations necessary for encoding and decoding quantum codes.  It is
also the group generated by fault-tolerant bitwise operations performed on
qubits that are encoded by certain quantum codes \cite{CaSh96},
\cite{ShoF}, \cite{SteR}.  Investigation of the connection between this
group and existing quantum codes has led us to a general construction for
such codes which allows us to generate many new examples.  The initial
results of this study were reported in \cite{CRSS96}.  However, it is very
hard to construct codes using the framework of \cite{CRSS96}.  In the
present paper we develop the theory to the point where it is possible to
apply standard techniques from classical coding theory to construct quantum
codes.  Some of the ideas in \cite{CRSS96} (although neither the
connections with the Clifford group nor with finite geometries or fields)
were discovered independently by Gottesman \cite{Got96}.

The paper is arranged as follows.  Section~2 transforms the problem into
one of constructing a particular type of binary space (Theorem~\ref{th1}).
Section~3 shows that these spaces in turn are equivalent to a certain class
of additive codes over $GF (4)$ (Theorem~\ref{th2}).  The rest of the paper
is then devoted to the study of such codes.  Their basic properties are
described in the remainder of Section~3, and Section~4 gives a number of
general constructions.  Sections~5, 6, and 7 then deal with cyclic and
related codes, self-dual codes, and bounds.  Until now little was known
about general bounds for quantum codes.  The linear programming bound
(Theorems~\ref{LPA} and \ref{LPW}) presented in Section~7 appears to give
quite sharp bounds for those codes.  This can be seen in the main table of
the paper, Table~III, given in Section~8, which is based on the results of
the earlier sections.  Although there are still a large number of gaps in
the table, the upper and lower bounds are generally not too far apart and
there are a considerable number of entries where the parameters of the best
codes are known exactly.  Section~9 contains an update on developments that
have occurred since the manuscript of this paper was first circulated.

In order to reduce the length of the paper, proofs which either use
standard techniques in coding theory or are straightforward will be
omitted.
\section{From quantum codes to binary spaces}
\hsp Recall from Section 1 that the quantum state space of $n$ qubits is
${\Bbb C}^{2^n}$.  The idea behind quantum error correction is to encode
quantum states into qubits so that errors or decoherence in a small number
of individual qubits will have little or no effect on the encoded data.
More precisely, an encoding of $k$ qubits into $n$ qubits is taken to be a
linear mapping of ${\Bbb C}^{2^k}$ onto a $2^k$-dimensional subspace of
${\Bbb C}^{2^n}$.  Since the error correction properties of this mapping
depend only on the subspace rather than on the mapping, the subspace itself
will be called the quantum error correcting code.

Correction of arbitrary errors in an arbitrary $2^k$-dimensional subspace
is in general infeasible, since errors which map states in the subspace to
other states in the subspace cannot be corrected (because the latter are
also permissible states).  To overcome this, we make use of the tensor
product decomposition of ${\Bbb C}^{2^n}$ into $n$ copies of ${\Bbb C}^2$.
Quantum error correcting codes are subspaces oriented so that any error in
a relatively small number of qubits moves the state in a direction
perpendicular to the coded subspace, and thus can be corrected.

A bit error in an individual qubit corresponds to applying the Pauli 
matrix
$\sigma_x = \left({0 \atop 1} ~ {1 \atop 0} \right)$
to that qubit, and a phase error to the Pauli matrix
$\sigma_z = \left( {1 \atop 0} ~ {0 \atop -1} \right)$.
The third Pauli matrix,
$\sigma_y = \left( {0 \atop i} ~ {-i \atop 0} \right) = i \sigma_x
\sigma_z$,
corresponds to a combination of bit and phase errors.  The group $E$ of
tensor products $\pm w_1 \otimes \cdots \otimes w_n$ and $\pm i w_1 \otimes
\cdots \otimes w_n$, where each $w_j$ is one of $I$, $\sigma_x , \sigma_y,
\sigma_z$, describes the possible errors in $n$ qubits.  $E$ is a subgroup
of the unitary group $U(2^n)$.  In general, there is a continuum of
possible errors in qubits, and there are errors in sets of qubits which
cannot be described by a product of errors in individual qubits.  For the
purposes of quantum error correction, however, we need consider only the
three types of errors $\sigma_x$, $\sigma_y$ and $\sigma_z$, since any
error-correcting code which corrects $t$ of these errors will be able to
correct arbitrary errors in $t$ qubits \cite{BDSW96}, \cite{EM96},
\cite{KL96}.  We do not go into the details of this result, but essentially
it follows from the fact that the matrices $I$, $\sigma_x$, $\sigma_y$ and
$\sigma_z$ form a basis for the space of all $2 \times 2$ matrices, and so
the tensor products of $t$ of these errors form a basis for the space of
$2^t \times 2^t$ matrices.

Our codes will thus be tailored for the error model in which each qubit
undergoes independent errors, and the three errors $\sigma_x$, $\sigma_y$
and $\sigma_z$ are all equally likely.  The results of \cite{BDSW96},
\cite{EM96}, \cite{KL96} show that any code which corrects these types of
quantum errors will be able to correct errors in arbitrary error models,
assuming the errors are not correlated among large numbers of qubits and
that the error rate is small.  For other error models it may be possible to
find codes which correct errors more efficiently than our codes do; this is
not discussed in this paper.

This section and Section~3 show how to convert the problem of finding
quantum-error-correcting codes into one of finding certain types of
classical error-correcting codes.  We do this in two stages.  The first
stage reduces the problem from a quantum (continuous) one to a classical
(discrete) problem in finite geometry.  The second stage converts the
latter to a coding theory problem.

The finite geometry problem can be summarized as follows.  Let $\oE$ denote
a $2n$-dimensional binary vector space, whose elements are written $(a|b)$
and which is equipped with the inner product
\beql{Eq2}
((a | b ) , ( a' | b' )) = a \cdot b' + a' \cdot b ~.
\eeq
This is a symplectic inner product, since it satisfies
$$((a|b), (a|b)) =0 ~.$$
Define the weight of $(a|b) = (a_1 \cdots a_n | b_1 \cdots b_n)$ to be the
number of coordinates $i$ such that at least one of $a_i$ and $b_i$ is 1.
The distance between two elements $(a|b)$, $(a' | b') \in \oE$ is defined
to be the weight of their difference.

Then we have the following theorem, which is an immediate consequence of
Theorem~1 of \cite{CRSS96}.
\begin{theorem}\label{th1}
Suppose $\oS$ is an $(n-k)$-dimensional linear
subspace of $\oE$ which is contained in its dual $\oS^\perp$ (with
respect to the inner product \eqn{Eq2}),
and is such that there are no vectors of weight $\le d-1$ in $\oS^\perp \setminus \oS$.
Then there is a quantum-error-correcting 
code mapping $k$ qubits to $n$ qubits which can correct $[(d-1)/2]$ errors.
\end{theorem}

We will describe such a quantum-error-correcting code by saying it has
parameters $[[n,k,d]]$, and call $d$ the {\em minimal distance} of the
code.  A code obtained via Theorem~\ref{th1} will be called an {\em
additive code}.  Almost all quantum-error-correcting codes known at the
present time are additive.  However, we will have occasion to discussion
more general codes in this paper, and will use the symbol $((n,K,d))$ to
indicate a code with minimal distance $d$ (see \cite{ShLa}) that encodes
$K$ states into $n$ qubits.  Of course, an $[[n,k,d]]$ code is also an
$((n, 2^k , d))$ code.

Readers who are most interested in the codes themselves could now proceed
directly to Section~3.

To motivate the following discussion we begin by describing classical
binary linear codes from a slightly unusual perspective.  A linear code $C$
is of course a linear subspace of $\ZZ_2^n$, where $\ZZ_2 = \{0,1\}$.  But
$\ZZ_2^n$ can also be regarded as the group of possible errors, i.e., $C$
is also a subgroup of the error group.  Furthermore, this subgroup $C$ has
the following characterization in terms of the error group: an error $e$ is
in $C$ precisely when translation by $e$ takes codewords to codewords and
thus cannot be detected.  $C$ corrects a set of errors if and only if the
sum of any two errors can be detected, i.e. lies outside $C$, except that
the sum may be the trivial error {\bf 0}, which, while it cannot be
detected, has no effect.

In the quantum setting, it is possible for a nontrivial error to be
undetectable and yet have no impact on the encoded state.  This suggests
that we should attempt to construct a quantum code from a pair of subgroups
of the quantum error group $E$.  One subgroup (which we will call $S'$)
specifies the undetectable errors, while the other (called $S$) is the
subgroup of $S'$ consisting of errors that have no effect on the encoded
state.  $S$ is the analogue of the zero subgroup in the classical coding
case.

It will turn out to be important to require that every element of $S'$ commutes with $S$.
This implies in particular that $S$ is abelian.
So we are led to consider when elements of $E$ commute.

The group\footnote{`$E$' stands for `error group', but also serves as a
reminder that $E$ is essentially an extraspecial 2-group.  The association
of extraspecial 2-groups with finite orthogonal spaces, underlying all of
this section, is a standard one in group theory (cf. \cite{Asch},
Theorem~23.10; \cite{Hupp}, Theorem~13.8).  We have made further use of
this theory in \cite{grass2}, \cite{grass3}.}  $E$ has order $2^{2n+2}$ and
center $\Xi (E) = \{ \pm I , \pm iI \}$.  The quotient group $\bar{E} = E/
\Xi (E)$ is an elementary abelian group of order $2^{2n}$, and hence a
binary vector space.  Let $V$ denote the vector space $\ZZ_2^n$, and label
the standard basis of $\CC^{2^n}$ by $| v \ra$, $v \in V$.  Every element
$e \in E$ can be written uniquely in the form
\beql{Eq1}
e =i^\la X(a) Z(b)
\eeq
where $\la \in \ZZ_4$, $X(a): | v \ra \to | v + a \ra$, $Z(b): | v \ra \to
(-1)^{b \cdot v} | v \ra$, for $a,b \in V$.  The element $X(a) Z(b)$
indicates that there are bit errors in the qubits for which $a_j =1$ and
phase errors in the qubits for which $b_j =1$.

If $e, e' \in E$ are given by \eqn{Eq1} then $ee' = \pm e'e$, where the
sign is $(-1)^{a \cdot b' + a' \cdot b}$.  This induces the symplectic
inner product given in \eqn{Eq2}:
$$((a|b), (a'|b')) = a \cdot b' + a' \cdot b ~,$$
where we write $(a|b)$ for the image of $X(a) Z(b)$ in $\bar{E}$.  Two
elements in $E$ commute if and only if their images in $\bar{E}$ are
orthogonal with respect to this inner product.

A subspace $\oS$ of $\oE$ is said to be {\em totally isotropic} if for all
$\os_1$, $\os_2 \in \oS$ the symplectic inner product $(\os_1 , \os_2 )
=0$.  A subgroup $S$ of $E$ is commutative if and only if its image $\oS$
in $\oE$ is totally isotropic.  The dimension of a totally isotropic
subspace is at most $n$.  The groups $X = \{ X(a) : a \in V \}$ and $Z = \{
Z(b) : b \in V \}$ are examples of subgroups of $E$ whose images $\oX$,
$\oZ$ have dimension $n$.

We define $S^\perp$ to be the lift of $( \bar{S})^\perp$ to $E$; or, in
other words, $S^\perp$ is the centralizer of $S$ in $E$.  We will take $S'$
to be $S^\perp$, that is, $S^\perp$ will be group of undetectable errors.

Since $S$ is abelian, its elements can be simultaneously diagonalized.
This induces a decomposition of $\CC^{2^n}$ into orthogonal eigenspaces.
In order for $S$ to act trivially on the code, it is necessary for the code
to lie entirely in one of these eigenspaces.  Since we also want $S^\perp$
to preserve the code, we take the code to be one of the eigenspaces, to be
denoted by $Q$ (say).  We call codes obtained in this way {\em additive}
codes.

To each eigenspace of $S$ there corresponds a homomorphism $\chi : S \to
{\Bbb C}$, under which each element of $S$ is mapped to the corresponding
eigenvalue.  Then $\chi$ is a character of $S$, and $\chi (iI) =i$.

Every element $e \in E$ normalizes $S$, and so conjugation by $e$ induces
an action on characters.  Since $S^\perp$ commutes with $S$, elements of
$S^\perp$ induce the trivial action on the characters.  Any element outside
$S^\perp$ negates the value of the character at each element of $S$ with
which it anticommutes.  In particular, it induces a nontrivial action on
the characters, and so $E/ S^\perp$ acts faithfully.

It follows that the orbit of any given character must have size $|E /
S^\perp |$.  If $\bar{S}$ has dimension $n-k$, $|E / S^\perp | =2^{n-k}$.
On the other hand there are $2^{n-k}$ characters of $S$ such that $\chi
(iI) =i$, since the quotient of any two such characters is a character of
$\bar{S}$.  Thus $E/ S^\perp$ acts transitively.  It follows that each
eigenspace must have the same dimension, namely $2^k$.

It remains to determine the error-correcting properties of the code $Q$.
In the classical setting, we can correct a set of errors when the quotient
(really, difference) of any pair of the errors lies outside the set $C
\setminus \{ 0 \}$, that is, can either be detected or acts trivially.
Analogously, we have the following lemma.
\begin{lemma}\label{Lem1}
An additive quantum-error-correcting code $Q$ with associated space
$\bar{S}$ can correct a set of errors $\Sigma \subseteq E$ precisely when
$\bar{e}_1^{-1} \bar{e}_2 \not\in \bar{S}^\perp \setminus \bar{S}$ for all
$e_1 , e_2 \in \Sigma$.
\end{lemma}
\paragraph{Proof.}
Suppose an error $e \in E$ has occurred.  In order to correct $e$ we must
find some error $e_1 \in E$ such that $e_1^{-1} e$ acts trivially on $Q$,
i.e., $e_1^{-1} e \in S$.  In other words, we must determine the coset $e
S$.  The hypothesis implies that every coset of $S^\perp$ contains at most
one coset of $S$ intersecting $\Sigma$.  It therefore suffices to determine
the coset $eS^\perp$.  Recall that $E / S^\perp$ permutes the eigenspaces
of $S$ regularly.  If we measure in which eigenspace we now lie (which we
can do because distinct eigenspaces are orthogonal) we can immediately read
off $eS^\perp$.  This measurement has no effect on the state, since the
state lies inside one of the eigenspaces.

On the other hand, suppose $e_1$ and $e_2$ are two errors such that
$\bar{e}_1^{-1} \bar{e}_2 \in \bar{S}^\perp \setminus \bar{S}$.  Any
correction procedure must take any state $e_1 (v) \in e_1 (Q)$ to $v$.
Since $e_1^{-1} e_2 \in S^\perp$, $e_2 (v) \in e_1 (Q)$, so $e_2 (v)$ is
corrected to $e_1^{-1} e_2 (v)$.  However, since $e_1^{-1} e_2 \not\in S$,
there is a state $v \in Q$ such that $e_1^{-1} e_2 (v)$ is not proportional
to $v$, and we have failed to correct $e_2$. \hfill $\Box$

It follows from the Lemma that if we let $d$ be the minimal weight of
$\bar{S}^\perp \setminus \bar{S}$, the code can correct the set of all
errors of weight at most $[(d-1)/2]$.  We have now completed the proof of
Theorem~\ref{Nth1}: $Q$ maps $k$ qubits into $n$ qubits and can correct
$[(d-1)/2]$ errors.

Recall that the eigenspaces of $S$ are in one-to-one correspondence with characters of $S$
satisfying $\chi (iI) =i$.
To determine which eigenspace contains a given state it is therefore enough to compute this character.
Moreover, since $\chi$ is a homomorphism, it suffices to compute the character on a basis for $\bar{S}$.
Each element of the basis thus provides one bit of information;
the collection of these bits is the {\em syndrome} of the error.
Of course, as in classical coding theory, identifying the most likely error given the syndrome can be a difficult problem.
(There is no theoretical difficulty, since in principle an
exhaustive search can always be used.)

{\em The Clifford groups}.
Encoding is carried out with the help of a family of groups
called Clifford groups.\footnote{We follow Bolt
et~al. ( \cite{SS4}, \cite{SS5}) in calling these Clifford groups.
The same name is used for a different family of groups by Chevalley \cite{Chev} and Jacobson \cite{Jac2}.}
There are both complex (denoted by $L$) and real (denoted by $L_R$) versions of these groups.

The {\em complex Clifford group} $L$ is defined to be the subgroup
of the normalizer of $E$ in $U(2^n)$ that
contains entries from ${\Bbb Q}[ \eta ]$, $\eta = (1+i) / \sqrt{2}$.
The full normalizer of $E$ in $U(2^n)$ has an infinite center consisting of
the elements $e^{2 \pi i \theta} I$, $\theta \in \RR$.
Although these central elements have no effect quantum-mechanically,
we wish to work with a finite group.
The smallest coefficient ring we can use is ${\Bbb Q}[ \eta ]$, since
$$\left\{ \frac{1}{\sqrt{2}} \left(
{1 \atop 1} ~ {1 \atop -1} \right) ~
\left( {1 \atop 0} ~ {0 \atop i} \right) \right\}^3 =
\left( {\eta \atop 0} ~ {0 \atop \eta} \right) ~.
$$

The {\em real Clifford group} $L_R$ is the real subgroup of $L$, or
equivalently the subgroup of $L$ with entries from ${\Bbb Q}[ \sqrt{2} ]$.
If we define $E_R$ to the real subgroup of $E$, then $L_R$ is the
normalizer of $E_R$ in the orthogonal group $O(2^n)$.  The group $E_R$
consists of the tensor products $\pm w_1 \otimes \cdots \otimes w_n$, where
each $w_j$ is one of $I$, $\sigma_x$, $\sigma_z$, $\sigma_x \sigma_z$.
$E_R$ is an extraspecial 2-group with order $2^{2n+1}$ and center $\{ \pm I
\}$, and $E_R/ \{ \pm I \} = E/ \Xi (E) = \bar{E}$.  For many applications
it is simpler to work with the real groups $E_R$ and $L_R$ rather than $E$
and $L$.

The following are explicit generators for these groups.
First, $L$ is generated by $E$,
all matrices of the form
\beql{Eq300}
I_2 \otimes \cdots \otimes I_2 \otimes H_2 \otimes I_2 \otimes \cdots \otimes I_2
~,
\eeq
where $I_2 = \left( {1 \atop 0} ~ {0 \atop 1} \right)$,
$H_2 = \frac{1}{\sqrt{2}} \left( {1 \atop 1} ~ {1 \atop -1} \right)$,
and all matrices ${\rm diag} (i^{\phi (v)} )_{v \in V}$, where $\phi$ is any $\ZZ_4$-valued quadratic form on $V$.
Similarly, $L_R$ is generated by $E_R$, \eqn{Eq300} and all matrices ${\rm diag} ((-1)^{\phi (v)} )_{v \in V}$, where $\phi$ is now any $\ZZ_2$-valued quadratic form on $V$.

We also record some further properties of $L$ and $L_R$:
\begin{itemize}
\item
$L/ \langle E, \eta I \rangle$ is isomorphic to the symplectic group
$Sp_{2n} (2)$
(the group of $2n \times 2n$ matrices over ${\Bbb Z}_2$ preserving the inner
product (1) \cite{ATLAS}).
\item
$L$ has order $8 | Sp_{2n} (2)| 2^{2n}$
$$= 2^{n^2 + 2n + 3}
\prod_{j=1}^n (4^j -1 ) ~.
$$
\item
$L_R/ E_R$ is isomorphic to the orthogonal group $O_{2n}^+ (2)$ \cite{ATLAS}.
\item
$L_R$ has order $2| O_{2n}^+ (2) | 2^{2n}$
$$= 2^{n^2+n+2} (2^n -1) \prod_{j=1}^{n-1} (4^j -1) ~.$$
\item
$L$ acts on $\bar{E}$ as the symplectic group $Sp_{2n} (2)$;
$L_R$ acts on $\bar{E}$ as the orthogonal group $O_{2n}^+ (2)$.
\end{itemize}

The groups $L$ and $L_R$ have arisen in several different contexts,
and provide a link between quantum codes, the Barnes-Wall lattices
\cite{SS4}, \cite{SS5}, \cite{SS22},
the construction of orthogonal spreads and Kerdock sets \cite{CCKS96}, the
construction of spherical codes \cite{Kaz}, \cite{Sid1},
\cite{Sid2}, and the construction of Grassmannian packings
\cite{grass2}, \cite{grass3}.
They have also occurred in several purely group-theoretic contexts ---
see \cite{CCKS96} for references.
These groups are discussed further in the final paragraphs
of the present paper.

{\em Encoding an additive code $Q$}.
Since $Sp_{2n} (2)$ acts transitively on isotropic subspaces, and $E$ acts
transitively on eigenspaces for a given subspace, the Clifford group $L$
acts transitively on additive codes.
One such code is the trivial code corresponding to the subspace
$\bar{S}$ with generators $(0 | e_i )$, $i=k+1 , \ldots, n$.
By transitivity we can find an element $\lambda \in L$ which takes the trivial code to $Q$.
Of course $\lambda$ is not unique.
Cleve and Gottesman \cite{CG96} have given explicit gate descriptions to doing this.

{\em Pure vs. degenerate}.
In the quantum coding literature there is an important
distinction made between degenerate and nondegenerate codes.
A {\em nondegenerate} code is one for which different elements of $E$ produce linearly
independent results when applied to elements of the code.
We will find it convenient to introduce a second dichotomy,
between pure and impure codes.
We will say that a code is {\em pure} if distinct elements of $E$ produce
orthogonal results.

It is straightforward to verify that, for additive codes, `pure' and
`nondegenerate' coincide.  In general, however, a pure code is
nondegenerate but the converse need not be true.

For many purposes the pure/impure distinction is the correct one to use for
generalizing results from additive to nonadditive codes, and we will
therefore use this terminology throughout the paper.

{\em Bases}.
To find an explicit basis for $Q$ we may proceed as follows.
Choose a maximal isotropic subspace $\bar{T}$ containing $\bar{S}$, and consider the 1-dimensional
eigenspaces of $T$.
We obtain a basis for $Q$ by selecting those eigenspaces for which the
corresponding character agrees with the given character on $S$.
(Equivalently, we may take all the eigenspaces lying inside $Q$.)
The choice of $T$ is of course not unique, and we have the same freedom in choosing a basis as we did earlier when choosing the element $\lambda$ of the Clifford group.

We conclude this section by restating Theorem~\ref{Nth1} in more detail.

\addtocounter{theorem}{-1}
\begin{theorem}\label{Nth1}
Suppose $\oS$ is an $n-k$-dimensional linear 
subspace of $\oE$ which is contained in its dual $\oS^\perp$ (with
respect to the inner product \eqn{Eq2}),
and is such that there are no vectors of weight $\le d-1$ in $\oS^\perp \setminus \oS$.
Then by taking an eigenspace
(for any chosen linear character) of $\oS$, we obtain a quantum-error-correcting code mapping $k$ qubits
to $n$ qubits which can correct $[(d-1)/2]$ errors.
\end{theorem}

\section{From binary spaces to codes over $GF(4)$}
\hsp
As is customary (cf.~\cite{MS77}) we take the Galois field $GF(4)$ to consist of the elements
$\{ 0,1, \om , \oom \}$, with $\om^2 = \om +1$, $\om^3 =1$, and conjugation defined by $\bar{x} = x^2$;
the trace map ${\rm Tr} : GF(4) \to \ZZ_2$ takes $x$ to $x + \bar{x}$.
The {\em Hamming weight} of a vector $u \in GF(4)^n$, written ${\rm wt} (u)$, is
the number of nonzero components, and the {\em Hamming distance} between
$u, u' \in GF (4)^n$ is
${\rm dist} (u, u') = {\rm wt}  (u-u')$.
The minimal Hamming distance between the members of a subset $C$ of
$GF(4)^n$ will be denoted by ${\rm dist} (C)$.

To each vector $v = (a|b) \in \oE$ 
we associate the vector $\phi (v) = \omega a + \oom b \in GF (4)^n$.
It is immediate that the weight of $v$ is equal to the Hamming weight of $\phi (v)$, and the distance between vectors $v = (a|b)$, $v' = (a' | b') \in \oE$ is equal to ${\rm dist} ( \phi (v) , \phi (v') )$.
The symplectic inner product of $v$ and $v'$ (see \eqn{Eq2}) is equal to
${\rm Tr} ( \phi (v) \cdot \overline{\phi (v')} )$,
where the bar denotes conjugation in $GF(4)$,
since
\begin{eqnarray*}
{\rm Tr} ( \phi (v) \cdot \overline{\phi (v')} ) & = & {\rm Tr} ((\om a + \oom b) \cdot ( \oom a' + \om b' )) \\
& = & (a \cdot a') {\rm Tr} (1) + (a \cdot b') {\rm Tr} ( \oom ) + (a' \cdot b) {\rm Tr} ( \om ) +
(b \cdot b') {\rm Tr} (1) \\
& = & a \cdot b' + a' \cdot b ~.
\end{eqnarray*}

If $\oS$ is a linear subspace of $\oE$ then $C = \phi ( \oS)$ is a subset
of $GF(4)^n$ which is closed under addition.
We shall refer to $C$ as an {\em additive} code over $GF(4)$,
and refer to it as an $(n, 2^k )$ code if it contains $2^k$ vectors.
If $C$ is also closed under multiplication by $\om$, we say it is {\em linear}.

The {\em trace inner product} of vectors $u,v \in GF (4)^n$ will be denoted by
\beql{Eq8}
u \ast v = {\rm Tr} \, u \cdot \ov =
\sum_{j=1}^n (u_j \ov_j + \ou_j v_j) ~.
\eeq

If $C$ is an $(n, 2^k )$ additive code, its dual is defined to be
\beql{Eq9}
C^\perp = \{ u \in GF(4)^n : u \ast v =0 ~~\mbox{for all}~~
v \in C \}~.
\eeq
Then $C^\perp$ is an $(n,2^{2n-k})$ code.
If $C \subseteq C^\perp$ we say $C$ is {\em self-orthogonal},
and if $C = C^\perp$ then $C$ is {\em self-dual}.

Theorem~\ref{th1} can now be reformulated.
\begin{theorem}\label{th2}
Suppose $C$ is an additive 
self-orthogonal subcode of $GF(4)^n$, containing $2^k$ vectors,
such that there are no vectors of weight $\le d-1$ in $C^\perp \setminus C$.
Then any eigenspace of $\phi^{-1} (C)$ is a quantum-error-correcting code with parameters $[[n,n-k, d]]$.
\end{theorem}

We say that $C$ is {\em pure} if there are no nonzero vectors of weight $< d$ in $C^\perp$;
otherwise we call $C$ {\em impure}.
Note that
the associated quantum-error-correcting code is pure in the sense of Section~2 if and only if $C$ is pure.
We also say that a quantum-error-correcting code is {\em linear} if the associated additive code $C$ is linear.

When studying $[[n,k,d]]$ codes we allow $k=0$, adopting the convention that this corresponds to a self-dual
$(n,2^n)$ code $C$ in which the minimal nonzero weight is $d$.
In other words, an $[[n,0,d]]$ code is ``pure'' by convention.
An $[[n,0,d]]$ code is then a quantum state such that, when subjected to a decoherence of $[(d-1)/2]$ coordinates, it is possible
to determine exactly which coordinates were decohered.
Such a code might be useful for example in testing whether certain storage locations for qubits are decohering faster than they should.
These codes are the subject of Section~6.

Most codes over $GF (4)$ that have been studied before this have been linear and duality has been defined with respect
to the hermitian inner product $u \cdot \ov$.
We shall refer to such codes as {\em classical}.
\begin{theorem}\label{MM1}
A linear code $C$ is self-orthogonal (with respect to the trace inner product \eqn{Eq8}) if and only if
it is classically self-orthogonal with respect to the hermitian inner product.
\end{theorem}
\paragraph{Proof.}
The condition is clearly sufficient.
Suppose $C$ is self-orthogonal.
For $u,v \in C$ let $u \cdot \ov = \af + \beta \om$, $\af , \beta \in \ZZ_2$.
Then ${\rm Tr}(u \cdot \ov ) =0$ implies $\beta =0$, and ${\rm Tr} (u \cdot \bar{\om}\bar{v} ) =0$ implies $\af =0$, so $u \cdot \ov =0$. \hfill $\Box$

\vsp
The following terminology applies generally to additive codes over $GF(4)$.
We specify an $(n,2^k)$ additive code by giving either a $k \times n$ {\em generator matrix}
whose rows span the code additively, or by listing the generators inside diamond brackets
$\langle~~\rangle$.
If the code is linear a $k/2 \times n$ generator matrix will
suffice, whose rows are a $GF(4)$-basis for the code.

Let $\sG_n$ denote the group of order $6^n n!$ generated by permutations of the $n$ coordinates,
multiplication of any coordinates by $\om$, and conjugation of any coordinates.
Equivalently, $\sG_n$ is the wreath product of $S_3$ by $S_n$ generated by permutations
of the coordinates and arbitrary permutations of the nonzero elements of $GF(4)$ in each coordinate.
$\sG_n$ preserves weights and trace inner products.
Two additive codes over $GF(4)$ of length $n$ are said to be {\em equivalent} if one can be obtained from the other by
applying an element of $\sG_n$.
The subgroup of $\sG_n$ fixing a code $C$ is its {\em automorphism group}
$Aut (C)$.
The number of codes equivalent to $C$ is then equal to
\beql{Eq10}
\frac{6^n n!}{Aut (C)} ~.
\eeq

We determine the automorphism group of an $(n, 2^k )$ additive code $C$ by the following
artifice.
We map $C$ to a $[3n,k]$ binary linear code $\beta (C)$ by applying the map $0 \to 000$, $1 \to 011$,
$\om \to 101$, $\oom \to 110$ to
each generator of $C$.
Let $\Omega$ denote the $(n,2^{2n} )$ code containing all vectors, and form $\beta ( \Omega )$.
Using a program such as MAGMA \cite{Mag1}, \cite{Mag2}, \cite{Mag3} we compute the automorphism groups of the binary linear code
$\beta (C)$ and $\beta ( \Omega )$;
their intersection is $Aut (C)$.

Any $(n,2^k )$ additive code is equivalent to one with generator matrix of the form
$$
\left[
\matrix{
I_{k_0} & \om B_1 & A_1 \cr
\om I_{k_0} & \om B_2 & A_2 \cr
0 & I_{k_1} & B_3 \cr
} \right] ~,
$$
where $I_r$ denotes an identity matrix of order $r$,
$A_j$ is an arbitrary matrix, $B_j$ is a binary matrix, and $k= 2k_0 + k_1$.
An $(n,2^k )$ code is called {\em even} if the weight of every codeword is even, and otherwise {\em odd}.
\begin{theorem}\label{M0}
An even additive code is self-orthogonal.
A self-orthogonal linear code is even.
\end{theorem}
\paragraph{Proof.}
The first assertion holds because
\beql{Eq11}
{\rm wt}  (u+v) \equiv {\rm wt}  (u) + {\rm wt}  (v) + u \ast v ~~ ( \bmod~2)
\eeq
for all $u,v \in GF(4)^n$, and the second because
\beql{Eq12}
u \ast (\om u) \equiv {\rm wt}  (u) ~~( \bmod~2) ~.
\eeq
~~ \hfill $\Box$

\vsp
The {\em weight distribution} of an $(n, 2^k)$ additive code $C$ is the sequence
$A_0, \ldots, A_n$, where $A_j$ is the number of vectors in $C$ of weight $j$.
It is easy to see that the weight distribution of any translate $u+C$, for $u \in C$, is the same as that of $C$,
and so the minimal distance between vectors of $C$ is equal to the minimal nonzero weight in $C$.
The polynomial $W(x,y) = \sum_{j=0}^n A_j x^{n-j} y^j$
is the {\em weight enumerator} of $C$ (cf. \cite{MS77}).
\begin{theorem}\label{M1}
If $C$ is an $(n,2^k)$ additive code with weight 
enumerator $W(x,y)$, 
the weight enumerator of the dual code $C^\perp$ is given by $2^{-k} W(x + 3y , x-y)$.
\end{theorem}
\paragraph{Proof.}
This result, analogous to the MacWilliams identity for linear codes, follows from
the general theory of additive codes developed by Delsarte \cite{Del72}, since our trace inner product
is a special case of the symmetric inner products used in \cite{Del72}. \hfill $\Box$

\section{General constructions}
\hsp
In this section we describe some general methods for modifying and combining additive codes over
$GF (4)$.

The {\em direct sum} of two additive codes is defined in the natural way:
$C \oplus C' = \{uv: u \in C , v \in C' \}$.
In this way we can
form the direct sum of two quantum-error-correcting codes, combining
$[[n,k,d]]$ and $[[n', k', d']]$ codes to produce an
$[[n+n', k+k' , d'' ]]$ code, where $d'' = \min \{ d, d'\}$.
An additive code which is not a direct sum is called {\em indecomposable}.
\begin{theorem}\label{P0}
Suppose an $[[n,k,d]]$ code exists.
(a)~If $k > 0$ then an $[[n+1, k,d]]$ code exists.
(b)~If the code is pure and $n \ge 2$ then an $[[n-1 , k+1, d-1]]$ code exists.
(c)~If $k > 1$ or if $k=1$ and the code is pure, then an $[[n, k-1, d]]$ code exists.
(d)~If $n \ge 2$ then an $[[n-1, k, d-1 ]]$ code exists.
(e)~If $n \ge 2$ and the associated code $C$ contains a vector of weight 1 then an $[[n-1,k,d]]$ code exists.
\end{theorem}
\paragraph{Proof.}
Let $C$ and $C^\perp$ be the associated $(n, 2^{n-k})$ and
$(n, 2^{n+k})$ additive codes, respectively, with $C \subset C^\perp$.
(a)~Form the direct sum of $C$ with $c_1 = \{0,1\}$.
The resulting $[[n+1,k,d]]$ code is impure (which is why the construction fails for $k=0$).
(b)~Puncture $C^\perp$ (cf. \cite{MS77}) by 
deleting the first coordinate, obtaining an $(n-1, 2^{n+k} )$ code
$B^\perp$ (say) with minimal distance at least $d-1$.
The dual of $B^\perp$ consists of the 
vectors $u$ such that $0u \in C$, and so is contained in $B^\perp$.
(c)~There are $(n, 2^{n-k+1})$ 
and $(n, 2^{n+k-1})$ additive 
codes $B$ and $B^\perp$ with $C \subset B \subset B^\perp \subset C^\perp$.
(d)~Take $B= \{ u: 0u ~\mbox{or}~1u \in C \}$, 
so that $B^\perp = \{ v: 0v ~\mbox{or}~1v \in C^\perp \}$.
The words in $B^\perp \setminus B$ arise from truncation of words in
$C^\perp\setminus C$.
Any words in $C^\perp \setminus C$ of weight less 
than $d$ either begin with $\om$ or $\oom$, and so are not in 
$B^\perp$, or begin with a 0 or 1, 
and so (after truncation) are in $B^\perp \setminus B$.
Words of weight $d$ in $C^\perp$ beginning with 1 become words of weight $d-1$, so the minimal distance
in general is reduced by 1.
The proof of (e) is left to the reader. \hfill $\Box$

\vsp
To illustrate
Part (a) of the theorem, from the $[[5,1,3 ]]$ Hamming code (see Section~5) we obtain an impure $[[6,1,3]]$ code.
On the other hand exhaustive search (or integer programming, see Section~7) shows that no pure $[[6,1,3]]$ exists.
This is the first occasion when an impure code exists but a pure one does not.

A second $[[6,1,3]]$ code, also impure not equivalent to the first, is generated by 000011, 011110, $0 \om \om \om \om \om$, $101 \om \oom \om$, $\om 0 \om \oom 1 0$.
Up to equivalence, there are no other $[[6,1,3]]$ codes.

If we have additional information about $C$ then there is a more powerful technique (than that in Theorem~\ref{P0}(d)) for shortening a code.
\begin{lemma}\label{P1}
Let $C$ be a linear self-orthogonal code over $GF(4)$.
Suppose $S$ is a set of coordinates of $C$ such that every codeword of $C$ meets $S$ in a vector of even weight.
Then the code obtained from $C$ by deleting the coordinates in $S$ is also self-orthogonal.
\end{lemma}
\paragraph{Proof.}
Follows from Theorem~\ref{M0}. \hfill $\Box$
\begin{theorem}\label{P2}
Suppose we have a linear $[[n,k,d]]$ code with associated $(n,2^{n-k} )$ code $C$.
Then there exists a linear $[[n-m, k', d' ]]$ code with $k' \ge k-m$ and $d' \ge d$, for any $m$ such that there exists a codeword
of weight $m$ in the dual of the binary code generated by the supports of the codewords of $C$.
\end{theorem}

\paragraph{Proof.}
Let $S$ be the support of such a word of weight $m$.
Then $S$ satisfies the conditions of the Lemma,
and deleting these coordinates gives the desired code. \hfill $\Box$

\vsp
For example, consider the $[[85,77,3]]$ Hamming code given in the following section.
The code $C$ is an $(85, 2^4)$ code, and the supports of the codewords in $C$ generate a binary code with weight enumerator
$$x^{85} + 3570 x^{53} y^{32} + 38080 x^{45} y^{40} + 
23800 x^{37} y^{48} + 85 x^{21} y^{64} ~.$$
The MacWilliams
transform of this (\cite{MS77}, Theorem~1, p.~127) shows that the dual binary code contains vectors of weights
0, 5 through 80, and 85.
 From Theorem~\ref{P2} we may deduce the existence of $[[9,1,3]]$, $[[10,2,3]] , \ldots , [[80,72,3]]$ codes
(see the entries labeled $S$ in the main table in Section~8).

There is an analogue of Theorem~\ref{P2} for additive codes, but the construction of the corresponding
binary code is somewhat more complicated.

The direct sum construction used inTheorem~\ref{P0}(a) can be generalized.
\begin{theorem}\label{P3}
Given two codes $[[n_1 , k_1, d_1 ]]$ and $[[n_2, k_2, d_2]]$ with $k_2 \le n_1$ we can construct
an $[[n_1 + n_2 - k_2, k_1, d]]$ code, where
$d \ge \min \{d_1, d_1+d_2 - k_2 \}$.
\end{theorem}

\paragraph{Proof.}
Consider the associated codes $C_1$, $C^\perp_1$ with parameters
$(n_1 , 2^{n_1 - k_1})$, $(n_1, 2^{n_1 + k_1} )$ and $C_2$, $C^\perp_2$ with parameters
$(n_2 , 2^{n_2 - k_2})$,
$(n_2, 2^{n_2 + k_2} )$.
Let $\rho$ be the composition of the natural map from $C^\perp_2$ to $C^\perp_2/C_2$ 
with any inner-product preserving map from $C^\perp_2/C_2$ to $GF(4)^{k_2}$.
Then we form a new code $C= \{ uv: v \in C^\perp_2, u \rho (v) \in C_1 \}$,
with
$C^\perp= \{ uv: v \in C^\perp_2 , u \rho (v) \in C^\perp_1 \}$.
If $\rho (v) \neq 0$, $v$ contributes at least $d_2$
to the weight of $uv$, but $u$ need have weight only $d_1 - k_2$.
If $\rho (v) =0$, and $uv \neq 0$, ${\rm wt}  (u) \ge d_1$. \hfill $\Box$

\vsp
Different choices for $\rho$ may produce inequivalent codes.
Choosing $\rho$ corresponds to choosing an encoding method for $C_2$.

For example, if the second code is the $[[1,0,1]]$ code with generator matrix $[1]$, the new code has
parameters $[[n_1 +1, k_1, d_1 ]]$,
as in Theorem~\ref{P0}(a).
A different $[[n_1+1,k,d_1 ]]$ code is obtained if we take the second code
to be the $[[2,1,1]]$ code with generator matrix $[11]$.
In particular, the second $[[6,1,3]]$ code mentioned above may be
obtained in this manner.

Theorem~\ref{P3} can be used to produce an analogue of concatenated codes in the quantum setting.
If $Q_1$ is an $[[nm,k ]]$ code such that the associated $(nm, 2^{nm+k} )$ code has minimal nonzero weight $d$
in each $m$-bit block, and $Q_2$ is an $[[n_2, m, d_2]]$ code, then encoding
each block of $Q_1$ using $Q_2$ (as in Theorem~\ref{P3}) produces an
$[[nn_2, k, dd_2 ]]$ {\em concatenated} code.

A particularly interesting example is obtained by concatenating the $[[5,1,3]]$ Hamming code
(see Section~5) with itself.
We take $Q_1 = Q_2$, and let the associated linear $(5, 2^4)$ code have generator matrix
$\left[ \matrix {0 & 1 & 1 & 1 & 1 \cr 1 & 0 & 1 & \om & \oom \cr} \right]$.
Then we obtain a
$[[25,1,9]]$ code for which the associated $(25, 2^{24} )$ and 
$(25, 2^{26} )$ linear codes have the generator
matrices shown in Fig.~1.
Although the Hamming code is pure, the concatenated code is not.
\begin{figure}[htb]
\caption{Generator matrices for a $(25, 2^{24})$ linear code (above the line) and its dual,
a $(25, 2^{26} )$ linear code (all rows), corresponding to a $[[25,1,9]]$ quantum code.}

$$
\left[
\begin{array}{c@{\,}c@{\,}c@{\,}c@{\,}c@{\,}c@{\,}c@{\,}c@{\,}c@{\,}c@{\,}c@{\,}c@{\,}c@{\,}c@{\,}c@{\,}c@{\,}c@{\,}c@{\,}c@{\,}c@{\,}c@{\,}c@{\,}c@{\,}c@{\,}c@{\,}}
0&0&0&0&0&0&0&0&0&0&0&0&0&0&0&0&0&0&0&0&0&1&1&1&1 \\
0&0&0&0&0&0&0&0&0&0&0&0&0&0&0&0&0&0&0&0&1&0&1&\om&\oom \\
0&0&0&0&0&0&0&0&0&0&0&0&0&0&0&0&1&1&1&1&0&0&0&0&0 \\
0&0&0&0&0&0&0&0&0&0&0&0&0&0&0&1&0&1&\om&\oom&0&0&0&0&0 \\
0&0&0&0&0&0&0&0&0&0&0&1&1&1&1&0&0&0&0&0&0&0&0&0&0 \\
0&0&0&0&0&0&0&0&0&0&1&0&1&\om&\oom&0&0&0&0&0&0&0&0&0&0 \\
0&0&0&0&0&0&1&1&1&1&0&0&0&0&0&0&0&0&0&0&0&0&0&0&0 \\
0&0&0&0&0&1&0&1&\om&\oom&0&0&0&0&0&0&0&0&0&0&0&0&0&0&0 \\
0&1&1&1&1&0&0&0&0&0&0&0&0&0&0&0&0&0&0&0&0&0&0&0&0 \\
1&0&1&\om&\oom&0&0&0&0&0&0&0&0&0&0&0&0&0&0&0&0&0&0&0&0 \\
0&0&0&0&0&0&0&1&\oom&\om&0&0&1&\oom&\om&0&0&1&\oom&\om&0&0&1&\oom&\om \\
0&0&1&\oom&\om&0&0&0&0&0&0&0&1&\oom&\om&0&0&\om&1&\oom&0&0&\oom&\om&1 \\ \hline
0&0&0&0&0&0&0&0&0&0&0&0&1&\oom&\om&0&0&\oom&\om&1&0&0&\om&1&\oom
\end{array}
\right]
$$
\end{figure}

The construction of quantum codes used in \cite{CaSh96} can 
be restated in the present terminology
(and slightly generalized):
\begin{theorem}\label{P4}
Let $C_1 \subseteq C_2$ be binary linear codes.
By taking $C= \om C_1 + \oom C_2^\perp$ in Theorem~\ref{th2} 
we obtain an $[[n,k_2 -k_1, d]]$ code, where $d= \min \{ {\rm dist} 
(C_2 \setminus C_1 )$, ${\rm dist} (C_1^\perp \setminus C_2^\perp ) \}$.
\end{theorem}
\paragraph{Proof.}
It is easily verified that $C$ is additive and 
that $C \subseteq C^\perp = \oom C_1^\perp + \om C_2$. \hfill $\Box$

\vsp
Another construction based on binary codes due to Gottesman \cite{Got96}
can be generalized as follows.
\begin{theorem}\label{P5}
Let $\sS_m$ be the classical binary simplex code of length
$n=2^m -1$, dimension $m$ and minimal distance $2^{m-1}$ (Chapter~14 of \cite{MS77}).
Let $f$ be any fixed-point-free
automorphism of $\sS_m$ and let $\sG_m$ be the $(2^m, 2^{m+2} )$
additive code generated by the vectors $u+ \om f(u)$, $u \in \sS_m$, with a $0$ appended, together with the vectors
$11 \ldots 1$, $\om \om \ldots \om$ of length $2^m$.
This yields a $[[2^m, 2^m -m-2, 3 ]]$ quantum code.
\end{theorem}

We omit the proof.

We can show that $\sG_m$ has the following properties (again, to save space, the proofs are omitted).
\begin{itemize}
\item[(i)]
For any choice of $f$,
$\sG_m$ has weight enumerator
$$x^{2^m} + 4(2^m -1) x^{2^{m-2}} y^{3.2^{m-2}} + 3y^{2^m} ~.$$
\item[(ii)]
The vectors of weight $2^m$ generate a subcode of dimension 2.
\item[(iii)]
Suppose $\sG_m$ is constructed using the automorphism $f$,
and $\sG'_m$ using $f'$.
Then $\sG'_m$ is equivalent to $\sG_m$ if and only if $f'$ is conjugate
under $Aut ( \sS_m )$ to one of
\beql{Eq900}
\{ f, 1-f, 1/f , 1-1/f,
1/(1-f), f/(1-f) \} ~.
\eeq
\item[(iv)] The automorphism group of $\sG_m$ has a normal subgroup $H$
which is a semidirect product of the centralizer of $f$ in $Aut ( \sS_m )$
with $\sS_m$, the index $[Aut ( \sG_m ) : H ]$ being the number of elements
of \eqn{Eq900} that are conjugate to $f$.
\item[(v)]
$\sG_m$ is linear precisely when $f$ satisfies $f^2 + f+1 =0$.
\end{itemize}

Before giving some examples, we remark that $Aut ( \sS_m )$ is isomorphic to the general linear group $GL_m (2)$, and conjugacy classes of $GL_m (2)$
are determined by their elementary divisors.
So the most convenient way to specify $f$ is by listing its elementary divisors.

For $m=3$, there is a unique choice for $f$, with elementary divisor $x^3 + x +1$,
and so there is a unique $\sG_3$, with parameters $[[8,3,3]]$.
Then $Aut ( \sG_3 )$ has order 168, and is a semidirect product of a cyclic group $C_3$ with the general affine group $GA_1 (8)$.

For $m=4$ there are three distinct codes $\sG_4$, with parameters $[[16,10,3]]$.
The corresponding elementary divisors for $f$ are:

(a) $x^2 + x+ 1$ (twice).
This produces a linear code, with $| Aut ( \sG_4 ) | = 17280$.
(In general the code is linear precisely when all the elementary divisors are equal to $x^2 + x+1$.)

(b) $(x^2 +x+1)^2$, with $|Aut ( \sG_4 )| = 1152$.

(c) $x^4 +x+1$, with $|Aut ( \sG_4 ) | = 480$.

For $m=5$ there are two distinct $\sG_5$ codes, with parameters
$[[32, 25,3]]$.
The corresponding elementary divisors are

(a) $x^3 +x+1$ and $x^2 +x+1$, with $|Aut ( \sG_5 )| = 2016$.

(b) $x^5 +x^2 +1$, with $|Aut ( \sG_5 ) | = 992$.

Gottesman \cite{Got96a} used just a single $f$, which he took to be (if $m$ is even)
$$
\left[
\matrix{
0 & 1 & 0 & 0 & \cdots & 0 \cr
0 & 0 & 1 & 0 & \cdots & 0 \cr
~ & \cdot & \cdot & \cdot & \cdots & ~ \cr
0 & 0 & 0 & 0 & \cdots & 1 \cr
1 & 1 & 1 & 1 & \cdots & 1 \cr
}
\right]
$$
while if $m$ is odd the first row is complemented.
Gottesman's codes correspond to those labeled (c) (for $m=4$) and (b) (for $m=5$).

The codes in Theorem~\ref{P5} can be extended.
\begin{theorem}\label{P6}
For $m \ge 2$, there exists an $[[n, n-m-2,3]]$ code, where $n$ is
$$\sum_{i=0}^{m/2} 2^{2i} ~~(m~\mbox{even} ), ~~
\sum_{i=1}^{(m-1)/2} 2^{2i+1} ~~(m~\mbox{odd} ) ~.$$
\end{theorem}
\paragraph{Sketch of proof.}
The corresponding $(n, 2^{m+2} )$ additive code $C$ (say) has weight enumerator
$$x^n + (2^{m+2} -1 ) x^{n-2^m} y^{2^m} ~~(m~\mbox{even})$$
or
$$x^n + (2^{m+2} -2^m) x^{n-2^m +2} y^{2^m -2} +
(2^m -1) x^{n-2^m} y^{2^m} ~~(m~ \mbox{odd} ) ~.$$
We take $C_2$ and $C_3$ to be the additive codes
corresponding to the $[[5,1,3]]$ and $[[8,3,3]]$ codes already mentioned.
For $m > 3$,
let $\sG_m$ be as in Theorem~\ref{P5}, and let $\sG'_m$ be the subcode consisting of the
weight $2^m$ codewords in $\sG_m$.
Finally, let $\phi$ be any isomorphism between $C_{m-2}$ and $\sG_m / \sG'_m$ (note that both are
spaces of dimension $m$).
Define a new code $C_m$ to consist of all vectors $v_1 v_2$, where $v_1 \in C_{m-2}$ and $\phi (v_1) = v_2 + \sG'_m$.
A simple counting argument verifies that $C_m$ has the claimed weight distribution.
By applying Theorem~\ref{M1} we find that $C_m^\perp$ has minimal distance 3. \hfill $\Box$

\vsp
Theorem~\ref{P6} was independently discovered by Gottesman \cite{Got96a}.

The resulting codes, like those constructed in Theorem~\ref{P5}, are pure and
additive but in general are not linear.
For even $m$ we obtain
the Hamming codes of Section~5 as well as nonlinear codes with the same parameters.
For odd $m$ we obtain $[[8,3,3]]$, $[[40,33,3]]$, $[[168,159,3]]$, $\ldots$ codes.
A generator matrix for the $(40, 2^7)$ additive code
corresponding to a $[[40,33,3]]$ code is shown in Fig.~2.
\begin{figure}[htb]
\caption{Generator matrix for $(40, 2^7 )$ additive code, producing
a $[[40,33,3]]$ quantum code.}

$$
\left[
\begin{array}{c@{}c@{}c@{}c@{}c@{}c@{}c@{}c@{}c@{}c@{}c@{}c@{}c@{}c@{}c@{}c@{}c@{}c@{}c@{}c@{}c@{}c@{}c@{}c@{}c@{}c@{}c@{}c@{}c@{}c@{}c@{}c@{}c@{}c@{}c@{}c@{}c@{}c@{}c@{}c@{}}
0&0&0&0&0&0&0&0&1&1&1&1&1&1&1&1&1&1&1&1&1&1&1&1&1&1&1&1&1&1&1&1&1&1&1&1&1&1&1&1 \\
0&0&0&0&0&0&0&0&\om&\om&\om&\om&\om&\om&\om&\om&\om&\om&\om&\om&\om&\om&\om&\om&\om&\om&\om&\om&\om&\om&\om&\om&\om&\om&\om&\om&\om&\om&\om&\om \\
0&0&1&\oom&\om&\om&\oom&1&\om&0&1&0&1&\oom&\om&\om&\oom&1&0&1&0&\om&\oom&1&0&\om&\oom&\om&\oom&1&0&0&1&\oom&\om&\oom&\om&0&1&\oom \\
0&1&\om&\om&1&0&\oom&\oom&0&\om&0&0&\om&0&\om&\oom&1&\oom&1&1&\oom&1&\oom&\oom&1&\oom&1&1&\oom&1&\oom&\om&0&\om&0&0&\om&0&\om&\om \\
0&\om&0&\oom&1&\oom&1&\om&0&1&1&0&0&1&1&\om&\om&\oom&\oom&\om&\om&\oom&\oom&0&0&1&1&0&0&1&1&\om&\om&\oom&\oom&\om&\om&\oom&\oom&0 \\
1&0&\om&0&\om&\oom&1&\oom&0&0&0&\om&\om&\om&\om&1&1&1&1&\oom&\oom&\oom&\oom&0&0&0&0&\om&\om&\om&\om&1&1&1&1&\oom&\oom&\oom&\oom&0 \\
\om&0&\om&1&\oom&1&\oom&0&0&0&0&1&1&1&1&0&0&0&0&1&1&1&1&\om&\om&\om&\om&\oom&\oom&\oom&\oom&\om&\om&\om&\om&\oom&\oom&\oom&\oom&0 \\
\end{array}
\right]
$$
\end{figure}

The ``$u|u+v$'' construction for binary codes (\cite{MS77}, page~76) has an
analogue for quantum codes.
\begin{theorem}\label{P7}
Suppose there is a pure $[[n,k_1,d_1]]$ code with 
associated $(n, 2^{n-k_1})$ additive code $C_1$,
and a pure $[[n,k_2, d_2]]$ code with associated code $C_2$,
such that $C_1 \subseteq C_2$.
Then there exists a pure $[[2n, k_1 - k_2, d]]$ code, where $d=  \min \{ 2d_1, \delta \}$, $\delta = {\rm dist} (C_2)$.
\end{theorem}
\paragraph{Proof.}
Take $C$ to be the $(2n, 2^{2n-k_1 +k_2})$ additive code consisting of the vectors $u|u+v$,
$u \in C_2^\perp$, $v \in C_1$, where the bar denotes concatenation.
Then $C^\perp = \{u| u+v : u \in C_1^\perp ,~ v \in C_2 \}$ has minimal distance $\min \{ 2d_1 , \delta \}$,
by Theorem~33 of \cite{MS77}, Chapter~1. \hfill $\Box$

\vsp
For example, by combining the $[[14,8,3]]$ and $[[14,0,6]]$ codes shown in
Table~II of the next section we obtain a $[[28,8,6]]$ code.

Concerning the structure of additive but nonlinear codes, it is pointless to simply add one
generator to a linear code.
For if $D$ is an $(n, 2^{n+k})$ linear code, 
and $D' = \langle D, v \ra$ is an $(n,2^{n+k+1})$
additive code with minimal distance $d$, then it is easy to show that the linear code $D'' = \langle D, v, \om v \ra$
also has minimal distance $d$.

We end this section by listing some trivial codes.
An $[[n,k,1]]$ code exists for all $0 \le k \le n$, $n \ge 1$.
An $[[n,k,2]]$ code exists provided $0 \le k \le n-2$,
if $n \ge 2$ is even, or provided $0 \le k \le n-3$ if $n \ge 3$ is odd.
\section{Cyclic and related codes}
\hsp
An $(n, 2^k )$ additive code $C$ is {\em constacyclic} if there is a constant $\kappa$
(which in our case will be $1, \om$ or $\oom$) such that
$(u_0, u_1, \ldots , u_{n-1}) \in C$ implies
$(\kappa u_{n-1} , u_0 , u_1, \ldots, u_{n-2} ) \in C$.
If $\kappa =1$ the code is {\em cyclic}.
Besides these standard terms from the classical theory, we also need a new concept:
if $(u_0, u_1, \ldots, u_{n-1} ) \in C$ implies
$(\ou_{n-1} , u_0 , u_1, \ldots, u_{n-2} ) \in C$, the code will be called
{\em conjucyclic}.

We begin with linear codes.
If vectors are represented by polynomials in the natural way, a {\em linear}
constacyclic code is represented by an ideal in the ring of polynomials
modulo $x^n - \kappa$ (\cite{MS77}, \cite{KP92}).
The latter is a principal ideal ring, so the code consists simply of all multiples of a
single generator polynomial $g(x)$, which must divide $x^n - \kappa$.
We assume $n$ is odd.
\begin{theorem}\label{C1}
A linear cyclic or constacyclic code with generator polynomial $g(x)$ is self-orthogonal if and only if
$$g(x) g^{\dagger} (x) \equiv 0 ~~(\bmod~ x^n - \kappa ) ~,$$
where if $g(x) = \sum_{j=0}^{n-1} g_j x^j$,
\beql{Eq13}
g^{\dagger} (x) =\kappa \bar{g}_0 +
\sum_{j=1}^{n-1} \bar{g}_{n-j} x^j ~.
\eeq
\end{theorem}

We omit the elementary proof (cf. \cite{CLP95}).
Note that
$$g^{\dagger} (x) \equiv \overline{g(x^{-1})} ~~(\bmod ~ x^n - \kappa ) ~.$$
The $\dagger$ operation induces an involution on factors of $x^n - \kappa$, so we can write
\beql{Eq14}
x^n - \kappa = \prod_i p_i (x) \prod_j (q_j (x) q_j^{\dagger} (x)) ~,
\eeq
where the $p_i$, $q_j$ and $q_j^{\dagger}$ are all distinct and
$p_i^{\dagger} = p_i$.
Then a divisor $g(x)$ of $x^n - \kappa$ 
generates a self-orthogonal linear constacyclic code if and only
if $g(x)$ is divisible by each of the $p_i$'s
and by at least one from each $q_j$, $q_j^{\dagger}$ pair.
\paragraph{Example.}
The classical {\em Hamming code}
$H$ over $GF(4)$ has length $n= (4^m -1)/3$, 
contains $4^{n-m}$ codewords and has minimal distance 3,
for $m \ge 1$
\cite{MS77}, \cite{Lin82}.
The dual code $C=H^\perp$ is a self-orthogonal 
linear code, and the corresponding
quantum code has parameters $[[n, n-2m , 3]]$, where $n= (4^m -1)/3$.
$C$ and $H$ are cyclic if $m$ is even, constacyclic if $m$ is odd.
For example when $m=2$ we can take $H$ to have 
generator polynomial $g(x) = x^2 + \om x +1$, a divisor of $x^5 -1$,
and when $m=3$ we take $g(x) = x^3 + x^2 + x + \om$, 
a divisor of $x^{21} - \om$.
These codes meet the sphere-packing bound \eqn{LP1}
(see Section~7) with equality.
The smallest Hamming code, a $[[5,1,3]]$ code, was 
independently discovered in the present
context by \cite{BDSW96} and \cite{LMPZ96}.
See also \cite{CRSS96}.

Hamming codes correct single errors.
In the classical theory the generalizations of Hamming 
codes that correct multiple errors are
known as BCH codes \cite{MS77}.
A similar generalization yields multiple-error correcting quantum codes.

Rather than giving a complete analysis of these codes, which 
involves a number of messy details,
we simply outline the construction and give some examples.
These {\em quantum BCH codes} may be cyclic or constacyclic.

In the cyclic case we let $\xi$ be a primitive $n$-th root of unity in some
extension field of $GF(4)$, and write each factor $q_j$ in \eqn{Eq14} as
$q_j (x) = \prod_{s \in S_j} (x- \xi^s )$, the {\em zero set} $S_j$ being a
cyclotomic coset modulo $n$ under multiplication by 4 (see \cite{MS77},
Chap.~7).  The zero set associated with $q_j^{\dagger}$ is then $-2S_j$.
We choose a minimal subset of the $q_j$'s subject to the conditions that
(a)~there is an arithmetic progression of length $d-1$ in the union of its
zero sets, for which the step size is relatively prime to $n$, and (b)~if
$q_j$ is chosen, $q_j^\dagger$ is not.  Let $B$ be the cyclic code whose
generator polynomial is the product of the $q_j$'s.  Then (a) guarantees
that $B$ has minimal distance at least $d$ and (b) guarantees that $B
\supset B^\perp$.  In this way we obtain a quantum error-correcting code
with parameters $[[n,k,d]]$, where $k= n-2 \deg g$.

A similar construction works in the constacyclic case, 
only now we choose $\xi$ to be a primitive $(3n)$-th root
of unity.

In the special case when $n= (4^m -1)/3$, most of 
the $q_j$ have degree $m$, and we obtain a sequence
of cyclic or constacyclic codes which provided $m$ is at least 4,
begins
$$[[n,n-2m,3]], [[n, n-4m, 4]], [[n,n-6m,5]],
[[n,n-8m,7 ]], \ldots~.
$$
For example when $m=4$ we obtain $[[85,77,3]]$,
$[[85,69,4]]$, $[[85,61,5]]$ and $[[85,53,7]]$ codes.

We now discuss additive (but not necessarily linear) codes.
Note that an additive constacyclic code 
(with $\kappa = \om$ or $\oom$) is necessarily linear.
\begin{theorem}\label{C2}
(a) Any $(n,2^k)$
additive cyclic code $C$ has two generators, and can be 
represented as $\langle \om p(x) + q(x) , r(x) \ra$,
where $p(x)$, $q(x)$, $r(x)$ are binary polynomials,
$p(x)$ and $r(x)$ divide $x^n -1$ $(\bmod~2)$,
$r(x)$ divides $q(x) (x^n -1) / p(x)$ $(\bmod~2)$, 
and $k=2n - \deg p - \deg r$.
(b)~If $\langle \om p' (x) + q' (x), r' (x) \ra$ is another 
such representation, then $p' (x) = p(x)$,
$r' (x) = r(x)$ and $q' (x) \equiv q(x)$ $(\bmod~ r(x))$.
(c)~$C$ is self-orthogonal if and only if
\begin{eqnarray*}
p(x) r(x^{n-1} ) & \equiv & p(x^{n-1}) r(x) 
\equiv 0 ~~(\bmod~x^n-1 ) ~, \\
p(x) q(x^{n-1}) & \equiv & p(x^{n-1} ) q(x) ~~(\bmod ~x^{n}-1 ) ~.
\end{eqnarray*}
\end{theorem}
\paragraph{Proof.}
(a)~Consider the map ${\rm Tr}: C \to \ZZ_2 [x] / (x^n -1)$ 
obtained by taking traces
\linebreak
componentwise.
The kernel of this map is a binary cyclic code, so 
can be represented uniquely as $\langle r(x) \ra$, where $r(x)$ divides
$x^n-1$.
The image of the map is similarly a binary cyclic 
code $\langle p(x) \ra$.
The original code is generated by $r(x)$ and some 
inverse image of $p(x)$, say $\om p (x) + q(x)$.
Finally, if $r(x)$ did not divide $q(x) (x^n -1) / p(x)$,
then $((x^n -1) / p(x)) ( \om p (x) + q(x))$ would be a 
binary vector of $C$ not in $\langle r(x) \ra$,
a contradiction.
We omit the proof of (b).
(c)~One readily verifies that the inner product of the 
vectors corresponding to $\omega f (x) + g(x)$ and 
$\omega h(x) + i(x)$ is given by the constant coefficient of
$$f(x) i (x^{n-1} ) + g(x) h(x^{n-1} ) ~~(\bmod~x^n -1 )~.$$
But then the inner product of the vectors corresponding 
to $\om f(x) + g(x)$ and $x^m ( \om h(x) + i(x))$ is given by the coefficient of $x^m$
in $f(x) i (x^{n-1} ) + g(x) h(x^{n-1})$.
The result follows immediately. \hfill $\Box$

\vsp
We remark without giving a proof that if $C$ is 
self-orthogonal we may assume that $q(x)$ satisfies
\beql{Eq15}
q(x^{n-1} ) = \frac{\pi (x)}{p(x)} +
\frac{ \sigma (x) (x^n-1)}{p(x)} ~,
\eeq
and $r(x)$ divides $q(x) (x^n -1) / p(x)$, where
$\pi (x) \equiv \pi (x^{n-1} )$ $(\bmod~x^n -1)$, 
$\pi (x) \equiv 0$ $(\bmod~p(x))$,
and $\deg \sigma < \deg r + \deg p-n$.
This makes it possible to search through all 
self-orthogonal additive cyclic
codes of a given dimension:
$r(x)$ ranges over all divisors of $x^n -1$,
$p(x)$ ranges over all divisors of 
$(x^n -1) / gcd \{ r(x^{n-1} ) , x^n -1 \}$ of the 
appropriate degree, and finally
all choices for $\pi (x)$ and $\sigma (x)$ must be considered.

Table~I lists some additive cyclic codes that were found in this way.
\begin{table}[htb]
\caption{Cyclic codes.}
$$
\begin{array}{cc@{}c@{}c@{}c@{}c@{}c@{}c@{}c@{}c@{}c@{}c@{}c@{}c@{}c@{}c@{}c@{}c@{}c@{}c@{}c@{}c@{}c@{}c@{}c@{}c@{}c@{}c@{}c@{}c@{}c@{}c@{}c@{}c@{}c@{}c@{}c@{}c@{}c@{}c@{}c@{}c@{}c@{}c@{}c@{}}
\mbox{\underline{Parameters}} & \multicolumn{43}{c}{\mbox{\underline{Generators for additive code}}} \\ [+.1in]
[[15,0,6]] & \om & 1 & 1 & 0 & 1 & 0 & 1 & 0 & 0 & 1 & 0 & 1 & 0 & 1 & 1 \\ [+.1in]
[[21,0,8]] & \oom & \oom & 1 & \om & 0 & 0 & 1 & 1 & 1 & 1 & 0 & 1 & 0 & 1 & 1 & 0 & 1 & 1 & 0 & 0 & 0 & , & 1 & 0 & 1 & 1 & 1 & 0 & 0 & 1 & 0 & 1 & 1 & 1 & 0 & 0 & 1 & 0 & 1 & 1 & 1 & 0 & 0 \\ [+.1in]
[[23,0,8]] & \om & 0 & 1 & 0 & 1 & 1 & 1 & 1 & 0 & 0 & 0 & 0 & 0 & 0 & 0 & 0 & 1 & 1 & 1 & 1 & 0 & 1 & 0 \\ [+.1in]
[[23, 12,4]] & \oom & \oom & \om & \oom & \om & 1 & 1 & \oom & 1 & 1 & \om & 1 & \om & 1 & 0 & 1 & 1 & 0 & 0 & 0 & 0 & 0 & 0 \\ [+.1in]
[[25,0,8]] & 1 & 1 & 1 & 0 & 1 & 0 & \om & 0 & 1 & 0 & 1 & 1 & 1 & 0 & 0 & 0 & 0 & 0 & 0 & 0 & 0 & 0 & 0 & 0 & 0
\end{array}
$$
\end{table}
\begin{theorem}\label{C3}
Let $C$ be an $(n, 2^k )$ additive conjucyclic code, and form the binary code
$$C' = \{ {\rm Tr} (\om u ) | {\rm Tr} ( \oom  u ) : u \in C \} ~,$$
when the trace is applied componentwise and the bar denotes concatenation.
Then $C'$ is a binary cyclic code of length $2n$, which 
is self-orthogonal if and only if
$C$ is self-orthogonal.
\end{theorem}

We omit the proof.
Note that $C'$ determines $C$, since
$$\om {\rm Tr} ( \om  u) + \oom {\rm Tr} ( \om u ) = u ~.$$
Theorem~\ref{C3} makes it possible to search for codes of this type.
So far no record codes have been found.

We now return to linear codes.
A {\em quasicyclic} code is a code of length $n= ab$ on which the 
group acts as $a$ cycles of length $b$.
T.~A. Gulliver of Carleton University (Canada) and the 
University of Canterbury (New Zealand) has extensively
studied quasicyclic codes over small fields \cite{GB92}.
The last five examples in Table~II were found by him.
Double parentheses indicate the permutation to be applied.
\begin{table}[htb]
\caption{Linear quasicyclic codes.}
$$
\begin{array}{cl}
\mbox{\underline{Parameters}} & 
\multicolumn{1}{c}{\mbox{\underline{Generator}}} \\  [+.1in]
[[14,0,6]] & ((1000000)) ~ ((\oom 1 \oom \om 00 \om )) \\ [+.1in]
[[14,8,3]] & ((1011100))~ ((1 \oom \om \om 1 0 \oom )) \\ [+.1in]
[[15,5,4]] & ((10000))~ ((11\oom 00))~ ((11\om  \om 0)) \\ [+.1in]
[[18,6,5]] & ((110000))~ ((101\oom 00))~((11\om 1\om 0)) \\ [+.1in]
[[20,10,4]] & ((10000))~((1\oom 100))~((1111\om))~((11\oom \om \oom)) 
\\ [+.1in]
[[25,15,4]] & ((1 0000))~((1 \om 1 \om 0))~ (0101 \oom ))~ (( 1  \om \oom \om 1))~((10\om \om 0)) \\ [+.1in]
[[28,14,5]] & ((\om \om \oom 1 000))~(( \oom 0 \oom 1 000))~ ((1 \oom \oom 1 \om \oom 0 ))~(( \oom \om \oom \om \om 00)) \\ [+.1in]
[[30,20,4]] & ((11100))~ (( 10 \om 0 0))~ ((1 1 \oom \om 0))~((1 \om 1 \om \oom ))~ ((1 0  \om 1 0))~((1 \om  1 0 0)) \\ [+.1in]
[[40,30,4]] & ((00 1 \om \om )) ~ ((011 \om 1 )) ~ ((0010 \oom ))~ ((001 \om 1)) ~(00101) ~((1 \om 1 \om \oom ))~ ((111 \oom \om ))~ ((01 \om 1 \oom)) \\ 
\end{array}
$$
\end{table}
\section{Self-dual codes}
\hsp
In this section we study $[[n, 0, d]]$ 
quantum-error-correcting codes and their associated
$(n, 2^n )$ self-dual codes $C$.
These codes are of interest in their own right ---
for instance, the unique $[[2,0,2]]$ code corresponds to the quantum state
$\frac{1}{\sqrt{2}} (| 01 \rangle - | 10 \rangle)$, that is, an EPR pair.
They are also important for constructing $[[n,k,d]]$ codes 
with $k > 0$, as we will see in Section~8.

We begin with some properties of weight enumerators of self-dual codes.
\begin{theorem}\label{M2}
(a)~The weight enumerator of a self-dual code is fixed under 
the transformation
\beql{Eq16}
{\rm replace}~ {\binom{x}{y}} ~~{\rm by}~~ \frac{1}{2} 
\left( \begin{array}{cr}
1 & 3 \\ 1 & -1 \end{array} \right) {\binom{x}{y}} ~,
\eeq
and is therefore a polynomial in $x+y$ and $x^2 + 3y^2$.
(b)~The minimal distance of a self-dual code of 
length $n$ is $\le [n/2]+1$.
\end{theorem}
\paragraph{Proof.}
(a)~\eqn{Eq16} follows from Theorem~\ref{M1}, 
and the proof of the second
assertion is parallel to that of Theorem~13 of \cite{MOSW78}.
(b)~Parallel to the proof of Corollary~3 of \cite{MS73}. \hfill $\Box$

(The result in (b) has since been improved --- see Section~9.)
\begin{theorem}\label{M3}
(a)~The weight enumerator of an even self-dual code is 
a polynomial in $x^2 + 3y^2$ and
$y^2 (x^2 - y^2)^2$.
(b)~The minimal distance of an even self-dual code 
of length $n$ is $\le 2 [n/6]+2$.
\end{theorem}
\paragraph{Proof.}
(a)~This is an immediate consequence of Theorem~13 of \cite{MOSW78}.
(b)~From Corollary~15 of \cite{MOSW78}. \hfill $\Box$

\vsp
In view of the importance of {\em doubly-even} self-dual codes 
in binary coding theory,
we also note the following result.
\begin{theorem}\label{M4}
If there is an integer constant $c > 1$ such that the weight 
of every vector in a self-dual
code is divisible by $c$, then $c=2$.
\end{theorem}
\paragraph{Proof.}
The proof of the Gleason-Prange theorem for classical self-dual codes 
as given in \cite{Slo79} applies unchanged. \hfill $\Box$

\vsp
It is possible to give a complete enumeration of all self-dual codes 
of modest length, following the methods of \cite{MOSW78} and
\cite{CPS79}.
\begin{theorem}\label{M5}
(a)~The total number of self-dual codes of length $n$ 
is $\prod_{j=1}^n (2^j +1)$.
\begin{flushleft}
(b)$~~~~~~~~~~~~~~~~~~~~~~~~~~~~\ds \displaystyle\frac{1}{| Aut (C)|} =
\displaystyle\frac{\prod_{j=1}^n (2^j +1)}{6^n n!} ~,$
\end{flushleft}
where the sum is over all inequivalent self-dual codes $C$ of length $n$.
\end{theorem}
\paragraph{Proof.}
(a)~Parallel to that of Theorem~19 of \cite{MOSW78}.
(b)~From (a) and \eqn{Eq10}. \hfill $\Box$

\vsp
Let $d_n$ be the $(n, 2^{n-1} )$ code spanned by all even-weight 
binary vectors of
length $n$, $n \ge 2$, and let $d_n^+ = \langle d_n, \om \om \ldots \om \ra$.
\begin{theorem}\label{M6}
Suppose $C$ is a self-orthogonal additive code, in which no
coordinate is identically zero, and which is generated by words of weight 2.
Then $C$ is equivalent to a direct sum
$d_2^+ \oplus \ldots \oplus d_2^+ \oplus d_i
\oplus d_j \oplus d_k \oplus \ldots$~,
$i$, $j$, $k \ge 2$.
\end{theorem}
\paragraph{Proof.}
Analogous to that of Theorem~4 of \cite{CPS79}. \hfill $\Box$

With the help of Theorems~\ref{M5} and \ref{M6} we find that the 
numbers $t_n$ (respectively $i_n$)
of inequivalent (respectively inequivalent indecomposable)
self-dual codes of length $n$ for $n \le 5$ are
$$\begin{array}{cccccc}
n & 1 & 2 & 3 & 4 & 5 \\
t_n & 1 & 2 & 3 & 6 & 11 \\
i_n & 1 & 1 & 1 & 2 & 4
\end{array}
$$
This enumeration could be extended to larger values of $n$ 
without too much difficulty.

The indecomposable codes mentioned in the above table are 
the trivial code $c_1$, the codes $d_n^+$ for $n \ge 2$, the
length 4 code
$\langle 1100, 0011, \om \om \om \om , 01 \om \oom \ra$, the length 5 codes
$$
\langle 11000, 00110 , 00101, 01 \om \om \om , \om \om 001 \rangle$$
and
$$\langle 11 000, 00110, 10101, \om \om 00 \om , 00 \om \om \om \ra~,$$
and a $(5, 2^5)$ $d=3$ code obtained from the hexacode (see Section~8) 
using Theorem~\ref{P0}.

We have also investigated the highest achievable minimal distance 
of any self-dual code of length $n$,
or equivalently of any $[[n,0,d]]$ quantum-error-correcting code.
The results are shown in the $k=0$ column of the main table (Section~8).
Of course in view of Theorem~\ref{P0}(c) this also gives bounds on the 
minimal distance of any pure $[[n,k,d]]$ code.

We see from that table that the bound in Theorem~\ref{M3} for even 
self-dual codes is met with
equality at lengths $2,4, \ldots, 22, 28$ and 30.
In all but one of those cases the code can be taken to be a classical 
self-dual linear code over
$GF(4)$.
The exception is at length 12,
where although no classical self-dual codes exists with minimal 
distance 6 \cite{CPS79},
there is an additive code.
This is the $(12, 2^{12} )$ $d=6$ additive code
having generator matrix
$$
\left[
\begin{array}{c@{}c@{}c@{}c@{}c@{}c@{}c@{}c@{}c@{}c@{}c@{}c@{}}
0 & 0 & 0 & 0 & 0 & 0 & 1 & 1 & 1 & 1 & 1 & 1 \\
0 & 0 & 0 & 0 & 0 & 0 & \om & \om & \om & \om & \om & \om \\
1 & 1 & 1 & 1 & 1 & 1 & 0 & 0 & 0 & 0 & 0 & 0 \\
\om & \om & \om & \om & \om & \om & 0 & 0 & 0 & 0 & 0 & 0 \\
0 & 0 & 0 & 1 & \om & \oom & 0 & 0 & 0 & 1 & \om & \oom \\
0 & 0 & 0 & \om & \oom & 1 & 0 & 0 & 0 & \om & \oom & 1 \\
1 & \oom & \om & 0 & 0 & 0 & 1 & \oom & \om & 0 & 0 & 0 \\
\om & 1 & \oom & 0 & 0 & 0 & \om & 1 & \oom & 0 & 0 & 0 \\
0 & 0 & 0 & 1 & \oom & \om & \om & \oom & 1 & 0 & 0 & 0 \\
0 & 0 & 0 & \om & 1 & \oom & 1 & \om & \oom & 0 & 0 & 0 \\
1 & \om & \oom & 0 & 0 & 0 & 0 & 0 & 0 & \oom & \om & 1 \\
\oom & 1 & \om & 0 & 0 & 0 & 0 & 0 & 0 & 1 & \oom & \om
\end{array}
\right]~,
$$
which we will call the {\em dodecacode}.
This code is equivalent to the cyclic code with generator
$\om 1 0 1 00 1 001 01$.
It has weight distribution 
$A_0 =1$, $A_6 =396$, $A_8 = 1485$, $A_{10} =1980$, $A_{12} = 234$,
and its automorphism group has order 648
and acts transitively on the coordinates.

There is an interesting open question concerning length 24.
There exists a $(24, 2^{24} )$ $d=8$
classical code over $GF(2)$, the Golay code, and at least 
two $(24, 3^{12} )$ $d=9$
classical codes over $GF(3)$,
all meeting the analogous bounds to Theorem~\ref{M3}(b) \cite{MS77}.
It is known \cite{LP90} that there is no $(24, 4^{12} )$ $d=10$ 
classical code over $GF(4)$,
but the possibility of a $(24, 2^{24})$ $d=10$ {\em additive} 
self-dual code remains open.
Linear programming shows that if such a code exists then it must be even.
However, all our attempts so far to construct this code have failed,
so it may not exist.
\section{Linear programming and other bounds}
\hsp
Gottesman \cite{Got96} showed that any nondegenerate $[[n,k, 2t+1]]$ code 
must satisfy the sphere-packing bound
\beql{LP1}
\sum_{j=0}^t 3^j {\binom{n}{j}} \le 2^{n-k} ~.
\eeq
Knill and Laflamme \cite{KL96} have shown that any
(pure or impure) code must satisfy the following version
of the Singleton bound (cf. \cite{MS77}):
\beql{LP1b}
n \geq 4e+k ,
\eeq
where $e = \lf (d-1)/2 \rf$ is the number of errors correctable by
the code.
In this section we first establish a linear programming bound 
which applies to all $[[n,k,d]]$ codes, and then give
a slightly stronger version of the Singleton bound for pure codes.

Suppose an $[[n,k,d]]$ code exists, let $C$ be the 
corresponding $(n, 2^{n-k} )$ code over $GF(4)$ and let $C^\perp$,
an $(n, 2^{n+k})$ code, be its dual (see Theorem~\ref{th2}).
Let $A_0 , \ldots, A_n$ and $A'_0 , \ldots, A'_n$ be the 
weight distributions of $C$ and $C^\perp$ respectively.

In view of Theorem~\ref{P0}(e), we may assume that $A_1 =0$.
(Only minor modifications to Theorem~\ref{LPA} are required 
if this assumption is not made.)

The Krawtchouk polynomials appropriate for studying a code of 
length $n$ over $GF(4)$ will be denoted by
$$P_j (x,n) = \sum_{s=0}^j (-1)^s 3^{j-s} {\binom{x}{s}} 
{\binom{n-x}{j-s}} ~,$$
for $j=0, \ldots, n$ (see Chapter~6 of \cite{MS77}).
\begin{theorem}\label{LPA}
If an $[[n,k,d]]$ quantum-error-correcting code exists such 
that the associated $(n, 2^{n-k})$ code $C$ contains
no vectors of weight 1, then there is a solution to the following 
set of linear equations and inequalities:
\begin{eqnarray}
\label{LP6}
&&A_0 = 1, A_1 =0, A_j \ge 0 ~(2 \le j \le n) ~, \\
\label{LP7}
&&A_0 + A_1 + \cdots + A_n = 2^{n-k} ~, \\
\label{LP9}
&&A'_j = \frac{1}{2^{n-k}} \sum_{r=0}^n P_j (r,n) A_r ~~(0 \le j \le n) ~, \\
\label{LP10}
&&A_j = A'_j ~(0 \le j \le d-1) , ~ A_j \le A'_j ~(d \le j \le n) ~, \\
\label{LP11}
&&\sum_{j \ge 0} A_{2j} = 2^{n - k-1} ~~\mbox{or}~~ 2^{n-k}~,
\end{eqnarray}
\beql{LP12}
\frac{1}{2^{n-k-1}} \sum_{r=0}^n P_j (2r,n) A_{2r} \ge A'_j ~(0 \le j \le n) ~.
\eeq
(If the second possibility obtains in \eqn{LP11}, \eqn{LP12} just says that
$2A'_j \ge A'_j$ and can be omitted.)
\end{theorem}
\paragraph{Proof.}
\eqn{LP9} is a consequence of Theorem~\ref{M1}, and
\eqn{LP10} follows from the facts that $C \subset C^\perp$ and any vectors
in $C^\perp$ of weights between 1 and $d-1$ inclusive must also be in $C$.
From \eqn{Eq11}, the even weight vectors in $C$ form an additive subcode $C'$,
which is either half or all of $C$;
\eqn{LP11} then follows.
If $C'$ is half of $C$, then $C' \subset C \subset C^\perp \subset (C')^\perp$, which
yields \eqn{LP12}.
The other constraints are clear. \hfill $\Box$

\vsp
A more compact statement of the linear programming bound may be obtained by
rephrasing Theorem~\ref{LPA} in terms of weight enumerators.
\begin{theorem}\label{LPW}
If an $[[n,k,d]]$ quantum-error-correcting code exists then there
are homogeneous polynomials $W(x,y)$,
$W^\perp (x,y)$ and $S(x,y)$ of degree $n$ such that the following conditions hold:
\begin{eqnarray}
\label{EqLPW1}
W(1,0) & = & W^\perp (1,0) =1 ~, \\
\label{EqLPW2}
W^\perp (x,y) & = & 2^k W \left( \frac{x+3y}{2} , ~ \frac{x-y}{2} \right) ~, \\
\label{EqLPW3}
S(x,y) & = & 2^k W \left( \frac{x+3y}{2} , ~ \frac{y-x}{2} \right) ~, \\
\label{EqLPW4}
W^\perp (1,y) - W(1,y) & = & O (y^d) ~,
\end{eqnarray}
and
\beql{EqLPW5}
W(x,y) , W^\perp (x,y) - W(x,y) , S(x,y) \ge 0 ~,
\eeq
where $P(x,y) \ge 0$ indicates that the coefficients of $P(x,y)$ are nonnegative.
\end{theorem}
\paragraph{Proof.}
Take $W(x,y)$ to be the weight enumerator of $C$ and $W^\perp (x,y)$ to be the weight enumerator of $C^\perp$.
$S(x,y)$ is the {\em shadow enumerator} (by analogy with \cite{shadow}) and is nonnegative by Eq.~\eqn{LP12}. \hfill $\Box$

\vsp
We have implemented Theorems~\ref{LPA} and \ref{LPW} on the computer in two different ways.

(i)~We attempt to minimize $A_1 + \cdots + A_{d-1}$ subject to \eqn{LP6}--\eqn{LP12} using an
optimization program such as CPLEX \cite{CPLEX} or CONOPT \cite{Dru94}.
The AMPL language \cite{FGK93} makes it easy to formulate such problems and to switch from one package to another.

If all goes well, the program either finds a solution (which may lead to additional discoveries about the
code, such as that there must exist a vector of a particular weight), or else reports that no feasible solution exists,
in which case we can conclude that no $[[n,k,d]]$ code exists.

Unfortunately, for values of $n$ around 30,
the coefficients may grow too large for the problems to be handled using double precision arithmetic,
and the results cannot be trusted.\footnote{It is hoped that the multiple precision
linear programming package being developed by David Applegate of Rice University will
soon remove this difficulty.}

(ii)~Alternatively, using a symbolic manipulation program such as MAPLE \cite{Char91}, we may ask directly if there is a
feasible solution to \eqn{LP6}--\eqn{LP12}
or to \eqn{EqLPW1}--\eqn{EqLPW5} (the latter being easier to implement).
Since the calculations are performed in exact arithmetic, the answers are (presumably) completely reliable.
On the other hand
the calculations are much slower than when floating point arithmetic is used.

Most of the upper bounds in the main table were independently calculated using both methods.

When investigating the possible existence of a pure $[[n,k,d]]$ code, we
may set $A_2$ through $A_{d-1}$ equal to $0$.  In all cases within the
range of table III below, this had no effect; that is, the LP bound for
pure codes was the same as that for impure codes.  We handle \eqn{LP11} by
running the problem twice, once for each choice of the right-hand side.

For example, using Theorem~\ref{LPA} we find that there are no $[[n,1,5]]$ codes of length $n \le 10$ for which
$C$ has $A_1 =0$.
From Theorem~\ref{P0} we conclude that no $[[n,1,5]]$ code of any type exists with $n \le 10$.
On the other hand an $[[11,1,5]]$ code does exist --- see the following section.

Additional constraints can be included in Theorem~\ref{LPA} to reflect special knowledge about particular codes,
or to attempt to narrow the range of a particular $A_i$.
Many variations are the basic argument are possible, as illustrated in the following examples.

(i)~No $[[13,0,6]]$ code exists.
Let $C$ be a $(13,2^{13})$ additive code with $d \ge 5$,
and let $C'$ be its even subcode.
The linear constraints in Theorem~\ref{LPA} enable us to express all the unknowns in terms of $A_5$ and $A_6$.
The condition that the weight distribution of $(C')^\perp$
be integral implies certain congruence conditions on $A_5$ and $A_6$, from which it is possible to
eliminate $A_6$.
The resulting congruence implies $A_5 \equiv 1$ $(\bmod~2)$.
In particular $A_5 \neq 0$, and so $d=5$.

(ii)~No $[[18, 12, 3]]$ code exists.
Consider the $(18,2^6)$ additive code $C$.
Linear programming shows that $C$ must contain a vector of weight 12, which without loss of generality we may take to be $u_0 = 0^6 1^{12}$.
We define the refined weight enumerator of $C$ with respect to $u_0$ to be
$$R_C (x_0, x_1, y_0, y_1, y_2) = \sum_{u \in C} x_0^{6-a(u)} x_1^{a(u)} y_0^{12- b(u) -c(u)} y_1^{b(u)} y_2^{c(u)} ~,$$
where $a(u)$
is the weight of $u$ in the first 6 coordinates, and $b(u)$ (resp. $c(u)$) is the number of 1's
(resp. $\om$'s or $\oom$'s) in $u$ in the last 12 coordinates.
The conditions on $C$ imply that $c(u) \equiv 0$ $(\bmod~2 )$,
$$(a(u+u_0) , b(u+u_0) , c(u+u_0)) = (a(u) , 12- b(u) - c(u) , c(u)) ~,$$
and
$$R_{C^\perp} = \frac{1}{|C|} R_C (x_0 + 3x_1 , x_0 - x_1 , y_0 + y_1 + 2y_2 , y_0 + y_1 - 2y_2 , y_0 - y_1 ) ~.$$
By applying linear programming, we find that the weight distribution of $C$
must be either $A_0 =1$, $A_{12}= 9$, $A_{14} = 54$ or $A_0 =1$, $A_{12} =1$, $A_{13} = 24$, $A_{14} =30$, $A_{15} = 8$.
In either case, adding
these constraints to the refined weight enumerator produces a linear program with no feasible
solution.

(iii)~Similar arguments eliminate the parameters
$[[7,0,4]]$, $[[15,4,5]]$, $[[15,7,4]]$, $[[16,8,4]]$,
$[[19,3,3]]$, $[[22,14,4]]$ and $[[25,0,10]]$.

In the remainder of this section we briefly discuss another version
of the Singleton bound (cf. \eqn{LP1b}):
\begin{theorem}\label{SB1}
If a pure $[[n,k,d]]$ code exists then $k \le n-2d+2$.
\end{theorem}
\paragraph{Proof.}
The associated code $C^\perp$ is then an additive $(n, 2^{n+k})$ code with minimal distance $d$.
From Theorem~15 of \cite{Del72}, we have
$$2^{n+k} \le 4^{n-d+1} ~,$$
which implies $k \le n-2d+2$. \hfill $\Box$

\vsp
If $d$ is odd this coincides with the Knill and Laflamme bound \eqn{LP1b},
but is slightly stronger if $d$ is even.

We have determined all codes that meet this bound --- these are analogues of the classical MDS codes
(cf. Chapter~11 of \cite{MS77}).
Since the results are somewhat disappointing we simply state the answer and omit the rather lengthy proof.
\begin{theorem}\label{SB2}
A pure $[[n,n-2d+2, d]]$ code has parameters $[[n,n,1]]$ $(n \ge 1)$,
$[[n,n-2,2]]$ $(n$ even $\ge 2)$,
$[[5,1,3]]$ or $[[6,0,4]]$.
Up to equivalence there is a unique code in each case.
\end{theorem}

\vsp
Even allowing $k=n-2d+1$ does not appear to lead to any new codes.
Further analysis shows that any pure $[[n, n-2d+1,d]]$ code has parameters $[[n,n-1,1]]$ $(n \ge 1)$,
$[[n,n-3,2]]$ $(n \ge 3)$,
$[[5,0,3]]$ or $[[8,3,3]]$.

\section{A table of quantum-error-correcting codes}
\hsp
Table~III, obtained by combining the best upper and lower bounds given in the previous sections,
shows our present state of knowledge about the highest minimal distance $d$ in any $[[n,k,d]]$ code of length
$n \le 30$.
\begin{center}
\begin{tabular}{c} \hline
~~~~~~Table III about here~~~~~~ \\ \hline
\end{tabular}
\end{center}
\subsection*{Notes on Table~III}
\hsp
When the exact value of $d$ is not known, the lower and upper bounds are separated by a dash.

All unmarked upper bounds in the table come from the linear programming bound of Theorem~\ref{LPA}.
(A few of these bounds can also be obtained from Eq.~\eqn{LP1} or from Theorem~\ref{M2}.)
Unmarked lower bounds are from Theorem~\ref{P0}.
Note in particular that,
except in the $k=0$ column, once a particular value of $d$ has been achieved, the same value holds for
all lower entries in the same column using Theorem~\ref{P0}(a).
\begin{itemize}
\item[$\alpha$.]
A code meeting this upper bound must be impure (this follows from integer programming by an argument similar to that used in Section~7 to show that no
$[[13,0,6]]$ code exists).
\item[$\beta$.]
A special upper bound given in Section~7.
These bounds do not apply to nonadditive codes, for which the upper bound
must  be increased by 1.
\item[$\gamma$.]
This is the unique other entry in the table (besides those marked `$\beta$')
where the known upper bound for nonadditive codes is different from the bound for additive codes: if we omit \eqn{LP12} (which says that the code is either odd or even) from the linear program, the bound increases by 1.
In all other entries in the table, condition \eqn{LP12} is superfluous.
However, we will be surprised if a $((19, 2^8, 5))$ nonadditive code exists.
\end{itemize}

Most of the following lower bounds
are specified by giving the associated $(n,2^{n-k})$ additive code.
\begin{itemize}
\item[a.]
The {\em hexacode}, a $(6, 2^6)$ $d=4$ classical code that can be taken to be the $GF(4)$ span of
$\langle 001111, 0101 \om \oom , 1001 \oom \om \ra$
(see Chapter~3 of \cite{SPLAG}).
$Aut(h_6) = 3.S_6$, of order 2160.
\item[b.]
A classical self-dual code over $GF(4)$ --- see \cite{MOSW78}, \cite{CPS79}.
\item[c.]
A cyclic code, see Table~I.
\item[d.]
A $[[25,1,9]]$ code obtained by concatenating the $[[5,1,3]]$ Hamming code with itself (Fig.~1 of Section~4).
\item[e.]
The dodecacode defined in Section~6.
\item[f.]
An $[[8,3,3]]$ code, discovered independently in \cite{CRSS96}, \cite{Got96} and \cite{Ste96}.
The $(8,2^5)$ additive code may be generated by vectors $((01 \om \om \oom 1 \oom )) 0$,
$11111111$, $\om \om \om \om \om \om \om \om$ (where the double parentheses mean that all
cyclic shifts of the enclosed string are to be used).
Exhaustive search shows that this code is unique.
Another version is obtained from Theorem~\ref{P5}.
The automorphism group has order 168, and is the semidirect product of a cyclic group of order 3 and the general
affine group $\{ x \to ax +b : a,b,x \in GF (8) , a \neq 0 \}$.
\item[g.]
A quasicyclic code found by T.~A. Gulliver --- see Table~II of Section~5.
\item[h.]
A Hamming code, see Section~5.
\item[i.]
Use the $(12, 2^8)$ and $(14, 2^8)$ linear codes with generator matrices
$$
\left[
\begin{array}{c@{\,}c@{\,}c@{\,}c@{\,}c@{\,}c@{\,}c@{\,}c@{\,}c@{\,}c@{\,}c@{\,}c@{\,}}
0 & 0 & 0 & 0 & 0 & 0 & 1 & 1 & 1 & 1 & 1 & 1 \\
0 & 0 & 1 & 1 & 1 & 1 & 0 & 0 & 1 & 1 & \om & \om \\
0 & 1 & 0 & 1 & \om & \oom & 0 & 1 & 0 & \om & 1 & \om \\
1 & 0 & 0 & 1 & \oom & \om & 0 & 1 & \om & 0 & \om & 1 \\
\end{array}
\right]
$$
and
$$
\left[
\begin{array}{c@{\,}c@{\,}c@{\,}c@{\,}c@{\,}c@{\,}c@{\,}c@{\,}c@{\,}c@{\,}c@{\,}c@{\,}c@{\,}c@{\,}}
0 & 0 & 0 & 0 & 0 & 0 & 1 & 1 & 1 & 1 & 1 & 1 & 1 & 1 \\
0 & 0 & 1 & 1 & 1 & 1 & 0 & 0 & 0 & 0 & 1 & 1 & 1 & 1 \\
0 & 1 & 0 & 1 & \om & \oom & 0 & 1 & \om & \oom & 0 & 1 & \om & \oom \\
1 & 0 & 0 & 1 & \oom & \om & 0 & 1 & \oom & \om & 1 & 0 & \om & \oom \\
\end{array}
\right]
$$
respectively.
Their automorphism groups have orders 720 and 8064, and both act
transitively on the coordinates.
The first of these can be obtained from the $u|u+v$ construction
(c.f. Theorem~\ref{P7}) applied to the unique $[[6,4,2]]$ and $[[6,0,4]]$
codes.
\item[j.]
A $[[17,9,4]]$ code, for which the corresponding $(17,2^8)$ $d=12$ code $C$ is a well-known
linear code, a two-weight code of class TF3 \cite{CK86}.
The columns of the generator matrix of $C$ represent the 17 points of an ovoid in $PG(3,4)$.
Both $C$ and $C^\perp$ are cyclic, a generator for $C^\perp$ being $1 \om 1 \om 1 0^{12}$.
The weight distribution of $C$ is $A_0 =1$,
$A_{12} = 204$, $A_{16} =51$, and its automorphism group has order 48960.
\item[s.]
By shortening one of the following codes using Theorem~\ref{P2} or its additive analogue:
the $[[21,15,3]]$ or $[[85,77,3]]$ Hamming codes (see Section~5),
the $[[32,25,3]]$ Gottesman code (Theorem~\ref{P5}), 
the $[[40,30,4]]$ code given in Table~II or $[[40,33,3]]$ code shown in Fig.~2.
\item[u.]
From the $u|u+v$ construction (see Theorem~\ref{P7}).
\item[v.]
The following $(17,2^6)$ code with trivial automorphism group found by random search:
$$
\left[
\begin{array}{c@{}c@{}c@{}c@{}c@{}c@{}c@{}c@{}c@{}c@{}c@{}c@{}c@{}c@{}c@{}c@{}c@{}}
0&0&1&0&\om&\oom&\om&\oom&1&1&\om&\oom&0&0&1&1&\oom \\
0&0&\om&1&0&\om&0&\oom&\oom&\oom&1&1&\om&\oom&\oom&1&1 \\
0&1&0&0&1&\om&1&\om&\oom&\oom&\oom&0&\oom&1&\om&0&\oom \\
0&\om&0&\om&\om&0&\oom&1&\oom&1&\om&\oom&\om&1&\om&\om&1 \\
1&0&0&\om&\oom&0&0&1&\om&\om&\oom&1&\oom&\om&0&\oom&1 \\
\om&0&0&1&\oom&\oom&\oom&0&\oom&0&\oom&1&0&1&1&\om&\oom \\
\end{array}
\right]
$$
\end{itemize}

Comparison of the table with the existing tables \cite{BrSl96} of classical codes over $GF(4)$
reveals a number of entries where it may be possible to improve the lower bound by the
use of linear codes.
For example, classical linear $[30,18,8]$ codes over $GF(4)$
certainly exist.
If such a code can be found which contains its dual, we would obtain a $[[30,6,8]]$ quantum code.
\clearpage
\begin{table}[htb]
\caption{Highest achievable minimal distance $d$ in any $[[n,k,d]]$ quantum-error-correcting code.
The symbols are explained in the text.}
$$
\begin{array}{r||c|c|c|c|c|c|c|c|}
n \setminus k & 0 & 1 & 2 & 3 & 4 & 5 & 6 & 7 \\ \hline
~ & ~ & ~ & ~ & ~ & ~ & ~ & ~ & ~ \\ [-.15in] \hline
3 & 2 & 1 & 1 & 1 & ~ & ~ & ~ & ~ \\
4 & 2 & 2 & 2 & 1 & 1 & ~ & ~ & ~ \\
5 &  3 & ^h 3 & 2 & 1 & 1 & 1 & ~ & ~ \\
6 & ^a 4 & 3^\alpha & 2 & 2 & 2 & 1 & 1 & ~ \\
7 & 3^\beta & ^s 3 & 2 & 2 & 2 & 1 & 1 & 1 \\
8 & ^b 4 & ^s 3 & ^s 3 & ^f 3 & 2 & 2 & 2 & 1 \\
9 & 4 & ^s 3 & ^s 3 & ^s 3 & 2 & 2 & 2 & 1 \\
10 & ^b 4 & 4 & 4 & ^s 3 & ^s 3 & 2 & 2 & 2 \\
11 & 5 & 5 & 4 & ^s 3 & ^s 3 & ^s 3 & 2 & 2 \\
12 & ^e 6 & 5^\alpha & 4 & 4 & ^i 4 & ^s 3 & ^s 3 & 2 \\
13 & 5^\beta & 5 & 4 & 4 & 4 & 3-4 & ^s 3 & ^s 3 \\
14 & ^b 6 & 5 & 4-5 & 4-5 & 4 & 4 & ^i 4 & ^s 3 \\
15 & ^c 6 & 5 & 5 & 5 & ^g 4^\beta & 4 & 4 & ^s 3^\beta \\
16 & ^b 6 & 6 & 6 & 5 & 4-5 & 4-5 & 4 & 3-4 \\
17 & 7 & 7 & 6 & 5-6 & 4-5 & 4-5 & 4-5 & 4 \\
18 & ^b 8 & 7 & 6 & 5-6 & 5-6 & 5 & ^g 5 & 4 \\
19 & 7-8 & 7 & 6 & 5-6 & 5-6 & 5-6 & 5 & 4-5 \\
20 & ^b 8 & 7 & 6-7 & 5-7 & 5-6 & 5-6 & 5-6 & 4-5 \\
21 & ^c 8 & 7 & 6-7 & 5-7 & 5-7 & 5-6 & 5-6 & 4-6^\alpha \\
22 & ^b 8 & 7-8 & 6-8 & 5-7 & 5-7 & 5-7 & 5-6 & 4-6 \\
23 & ^c 8-9 & 7-9 & 6-8 & 5-8 & 5-7 & 5-7 & 5-7 & 5-6 \\
24 & ^b 8-10 & 8-9^\alpha & 6-8 & 6-8 & 6-8 & 6-7 & 6-7 & 5-7 \\
25 & ^c 8-9^\beta & ^d 9 & 7-8 & 7-8 & 7-8 & 7-8 & 6-7 & 5-7 \\
26 & 8-10 & 9 & 8-9 & 8-9 & 8 & 7-8 & 6-8 & 5-8 \\
27 & 9-10 & 9 & 9 & 9 & 8-9 & 7-8 & 6-8 & 5-8 \\
28 & 10 & 10 & 10 & 9 & 8-9 & 7-9 & 6-8 & 6-8 \\
29 & 11 & 11 & 10 & 9-10 & 8-9 & 7-9 & 6-9 & 6-8 \\
30 & ^b 12 & 11^\alpha & 10 & 9-10 & 8-10 & 7-9 & 6-9 & 6-9
\end{array}
$$
\end{table}
\clearpage
\centerline{Table~III cont.}
$$
\begin{array}{r||c|c|c|c|c|c|c|c|} 
n \setminus k & 8 & 9 & 10 & 11 & 12 & 13 & 14 & 15 \\ \hline
~ & ~ & ~ & ~ & ~ & ~ & ~ & ~ & ~ \\ [-.15in] \hline
3 & ~ & ~ & ~ & ~ & ~ & ~ & ~ & ~ \\
4 & ~ & ~ & ~ & ~ & ~ & ~ & ~ & ~ \\
5 & ~ & ~ & ~ & ~ & ~ & ~ & ~ & ~ \\
6 & ~ & ~ & ~ & ~ & ~ & ~ & ~ & ~ \\
7 & ~ & ~ & ~ & ~ & ~ & ~ & ~ & ~ \\
8 & 1 & ~ & ~ & ~ & ~ & ~ & ~ & ~ \\
9 & 1 & 1 & ~ & ~ & ~ & ~ & ~ & ~ \\
10 & 2 & 1 & 1 & ~ & ~ & ~ & ~ & ~ \\
11 & 2 & 1 & 1 & 1 & ~ & ~ & ~ & ~ \\
12 & 2 & 2 & 2 & 1 & 1 & ~ & ~ & ~ \\
13 & 2 & 2 & 2 & 1 & 1 & 1 & ~ & ~ \\
14 & ^s 3 & 2 & 2 & 2 & 2 & 1 & 1 & ~ \\
15 & ^s 3 & ^s 3 & 2 & 2 & 2 & 1 & 1 & 1 \\
16 & ^s 3^\beta & ^s 3 & ^s 3 & 2 & 2 & 2 & 2 & 1 \\
17 & 4 & ^j 4 & ^s 3 & ^v 3 & 2 & 2 & 2 & 1 \\
18 & 4 & 4 & ^s 3 & ^s 3 & 2^\beta & 2 & 2 & 2 \\
19 & 4^\gamma & 4 & 3-4 & ^s 3 & ^s 3 & 2^\beta & 2 & 2 \\
20 & 4-5 & 4 & ^g 4 & 3-4 & ^s 3 & ^s 3 & 2 & 2 \\
21 & 4-5 & 4-5 & 4 & ^s 4 & 3-4 & ^s 3 & ^s 3 & ^h 3 \\
22 & 4-6 & 4-5 & 4-5 & 4 & ^s 4 & 3-4 & ^s 3^\beta & ^s 3 \\
23 & 4-6 & 4-6 & 4-5 & 4-5 & ^c 4 & ^s 4 & 3-4 & ^s 3 \\
24 & 4-6 & 4-6 & 4-6 & 4-5 & 4-5 & 4 & ^s 4 & 3-4 \\
25 & 4-7 & 4-6 & 4-6 & 4-6 & 4-5 & 4-5 & 4 & ^g 4 \\
26 & 4-7 & 4-7 & 4-6 & 4-6 & 4-6 & 4-5 & 4-5 & 4 \\
27 & 4-8 & 5-7 & 4-7 & 4-6 & 4-6 & 4-5 & 4-5 & 4-5 \\
28 & ^u 6-8 & 5-8 & 5-7 & 5-7 & 5-6 & 5-6 & ^g 5-6 & 4-5 \\
29 & 6-8 & 5-8 & 5-7 & 5-7 & 5-6 & 5-6 & 5-6 & 4-5 \\
30 & 6-8 & 5-8 & 5-8 & 5-7 & 5-7 & 5-6 & 5-6 & 4-6
\end{array}
$$
\clearpage
\centerline{Table~III cont.}

$$
\begin{array}{r||c|c|c|c|c|c|c|c|} 
n \setminus k & 16 & 17 & 18 & 19 & 20 & 21 & 22 & 23 \\ \hline
~ & ~ & ~ & ~ & ~ & ~ & ~ & ~ & ~ \\ [-.15in] \hline
3 & ~ & ~ & ~ & ~ & ~ & ~ & ~ & ~ \\
4 & ~ & ~ & ~ & ~ & ~ & ~ & ~ & ~ \\
5 & ~ & ~ & ~ & ~ & ~ & ~ & ~ & ~ \\
6 & ~ & ~ & ~ & ~ & ~ & ~ & ~ & ~ \\
7 & ~ & ~ & ~ & ~ & ~ & ~ & ~ & ~ \\
8 & ~ & ~ & ~ & ~ & ~ & ~ & ~ & ~ \\
9 & ~ & ~ & ~ & ~ & ~ & ~ & ~ & ~ \\
10 & ~ & ~ & ~ & ~ & ~ & ~ & ~ & ~ \\
11 & ~ & ~ & ~ & ~ & ~ & ~ & ~ & ~ \\
12 & ~ & ~ & ~ & ~ & ~ & ~ & ~ & ~ \\
13 & ~ & ~ & ~ & ~ & ~ & ~ & ~ & ~ \\
14 & ~ & ~ & ~ & ~ & ~ & ~ & ~ & ~ \\
15 & ~ & ~ & ~ & ~ & ~ & ~ & ~ & ~ \\
15 & ~ & ~ & ~ & ~ & ~ & ~ & ~ & ~ \\
16 & 1 & ~ & ~ & ~ & ~ & ~ & ~ & ~ \\
17 & 1 & 1 & ~ & ~ & ~ & ~ & ~ & ~ \\
18 & 2 & 1 & 1 & ~ & ~ & ~ & ~ & ~ \\
19 & 2 & 1 & 1 & 1 & ~ & ~ & ~ & ~ \\
20 & 2 & 2 & 2 & 1 & 1 & ~ & ~ & ~ \\
21 & 2 & 2 & 2 & 1 & 1 & 1 & ~ & ~ \\
22 & 2 & 2 & 2 & 2 & 2 & 1 & 1 & ~ \\
23 & ^s 3 & 2 & 2 & 2 & 2 & 1 & 1 & 1 \\
24 & ^s 3 & ^s 3 & 2 & 2 & 2 & 2 & 2 & 1 \\
25 & 3-4 & ^s 3 & ^s 3 & 2 & 2 & 2 & 2 & 1 \\
26 & ^s 4 & 3-4 & ^s 3 & ^s 3 & 2 & 2 & 2 & 2 \\
27 & 4 & ^s 4 & 3-4 & ^s 3 & ^s 3 & 2 & 2 & 2 \\
28 & 4 & 4 & ^s 4 & 3-4 & ^s 3 & ^s 3 & 2 & 2 \\
29 & 4-5 & 4 & 4 & 3-4 & 3-4 & ^s 3 & ^s 3 & 2 \\
30 & 4-5 & 4-5 & 4 & 4 & ^g 4 & 3-4 & ^s 3 & ^s 3 
\end{array}
$$
\clearpage
\section{Subsequent developments}
\hsp
In the year and a half since the manuscripts of \cite{CRSS96} and the present
paper were first circulated there have been a number of further
developments.

(i) While we showed in Section~2 that the Clifford group $L$ suffices to encode additive codes, we did not give explicit
recipes for doing so.
Such recipes can now be found in Cleve and Gottesman \cite{CG96}.

(ii) The Cleve and Gottesman technique applies only to real (not complex) codes.
However, it can be shown \cite{RainsQSE} that any additive
code is equivalent to a real additive code (and any linear code is equivalent to a real linear code), so this is not a severe restriction.

(iii) DiVincenzo and Shor \cite{DiVS} have shown how to correct errors in additive
codes even when using imperfect computational gates.
The techniques of Shor \cite{ShoF} for performing computations on encoded
qubits using imperfect gates have been extended to general additive
codes by Gottesman \cite{GottFT}.

However, the most efficient methods currently known for fault-tolerant
computation \cite{AhB}, \cite{Kit97}, \cite{KLZ}, \cite{Ste97} use only
Calderbank-Shor-Steane codes (cf. Theorem~9).

(iv) It turns out that the proofs of the lower bounds on the capacity of quantum
channels given in Bennett et~al. \cite{BBP96}, \cite{BDSW96} and DiVincenzo, Shor and Smolin \cite{DiVSS} can be restated in terms of additive
codes.
In particular, this implies that these bounds can be attained using
additive codes.

(v) Cleve \cite{Cleve96} has found a way to apply asymptotic upper bounds for classical binary codes to additive codes.

(vi) Steane \cite{SteQR} has extended Gottesman's \cite{Got96} construction
(compare Theorem~\ref{P5}) to obtain quantum analogues of Reed-Muller codes.
The smallest of these new codes has parameters $[[32,10,6]]$.

(vii) The upper bounds in the column headed `$k=0$' in Table~III (with the
exception of the entries marked `$\beta$') have an obvious pattern with
period 6.  Further investigation of this pattern has led to an $n/3$ bound
for quantum codes (cf. Theorem~\ref{M3}) \cite{RainsQSE} and an analogous
$n/6$ bound for classical singly-even binary self-dual codes
\cite{RainsSB}.

(viii) The main construction in this paper (described in Section~2) can be generalized to primes greater than 2.
Some preliminary work along these lines has been done in
\cite{AhB}, \cite{Knill96}, \cite{Knill96a}, \cite{RainsNB}.

(ix) There are analogues of parts (a)--(c) of Theorem~\ref{P0} for nonadditive codes.
Parts (a) and (c) are trivial, while (b) now asserts that if a pure $((n,K,d))$ codes exists with $n \ge 2$ then an $((n-1, 2K, d-1))$ code exists \cite{RainsQWE}.

(x) How much of a restriction is it to use only additive quantum-error-correcting codes?
We conjecture: Not much!
So far essentially only one good nonadditive code has been
found.
This is the $((5,6,2))$ code described in \cite{RHSS}.
The best comparable additive code is a $((5,4,2))$ code.
The $((5,6,2))$ code can be used to construct a family of nonadditive
codes with parameters $((2m+1, 3.2^{2m-3}, 2))$ for all $m \ge 2$
\cite{RainsQd2}.
The $((5,6,2))$ code is optimal in that there exists no $((5,7,2))$ code.
It is not known if this is true for other codes in the family.
The next candidate for a good nonadditive code is at length 7, where we have unsuccessfully
tried to find a $((7,1,4))$ code.

(xi) Most of the upper bounds in this paper have only been proved to hold
for additive codes.
It turns out however that our strongest technique, the linear programming
bound of Theorem~\ref{LPW}, applies even to nonadditive codes with the appropriate definitions of $W$, $W^\perp$ (see \cite{ShLa}) and $S$ (see \cite{RainsQSE}).
The sole change needed in the statement of Theorem~\ref{LPW} is that $2^k$
must be replaced by $K$.

As a consequence, all but ten of the upper bounds in Table~III (those marked `$\beta$' or `$\gamma$') apply equally to nonadditive codes.

(xii) The purity conjecture.
As we have already remarked, in the range of Table~III the linear
programming bound for pure codes is no stronger than that for
impure codes.
Moreover, for several entries in the table a code meeting the linear programming
bound must be pure.
This suggests the following conjecture.
\paragraph{Conjecture.}
Let $K$ be the largest number (not necessarily an integer) greater than 1 such that there
exist polynomials $W$, $W^\perp$, $S$ as in the nonadditive version of
Theorem~\ref{LPW}.
Then for any such solution,
$$W(1,y) = 1 + O (y^d) ~,$$
or in other words the weight enumerator is pure.

This conjecture, together with some sort of monotonicity result about solutions to Theorem~\ref{LPW}, would imply the equivalence of the pure and impure linear programming bounds for
general (additive or nonadditive) codes.

We have verified the conjecture for all $n \le 50$.

(xiii)
Referring to the above conjecture, cases in which the extremal $K$ are powers of 2 are of particular interest.
In the range $n \le 45$ these are listed in the following table.

\begin{table}[htb]
\caption{Putative extremal quantum-error-correcting codes $((n,K,d))$ in which $K$ is a power of 2.}
$$
\begin{array}{ll}
\multicolumn{2}{l}{\mbox{(a) $K=2$:}} \\ 
((5,2,3)) & \mbox{(exists: Hamming code)} \\
((11,2,5)) & \mbox{(exists from dodecacode)} \\
((17,2,7)) & \mbox{(exists)} \\
((23,2,9)) & \mbox{(?)} \\
((29,2,11)) & \mbox{(exists: quadratic residue code)} \\
((35,2,13)) & \mbox{(?)} \\
((41,2,15)) & \mbox{(?)} \\ [+.2in]
\multicolumn{2}{l}{\mbox{(b) Two infinite families:}} \\
((2m,2^{2m-2},2)), m\ge 1 & \mbox{(exist)}\\
((n,2^{n-2m},3)), n=(4^m-1)/3, m \ge 2 & \mbox{(exist: Hamming codes)} \\ [+.2in]
\multicolumn{2}{l}{\mbox{(c) Some apparently sporadic possibilities:}} \\
((18,4096,3)) & \mbox{(?, must be nonadditive)} \\
((16, 256,4)) & \mbox{(?, must be nonadditive)} \\
((17, 512, 4)) & \mbox{(exists)} \\
((22, 2^{14}, 4)) & \mbox{(?, must be nonadditive)} \\
((27, 2^{15}, 5)) & \mbox{(?)} \\
((28, 2^{14}, 6)) & \mbox{(?)} \\
((40, 64, 13)) & \mbox{(?)} \\
\end{array}
$$
\end{table}

There are also some candidates for which $K$ is not a power of 2.
The first of these is $((5,6,2))$, and as mentioned above we were able to find such a code.
There is an infinite family of other candidates with $d=2$, none of which can exist \cite{RainsQd2}.
The remaining possibilities for $n \le 45$ are:
$$
\begin{array}{l}
((10, 24, 3)) \\
((13,40, 4)) \\
((21, 7168, 4 )) \\
((24, 49152, 4 )) \\
((22, 384, 6)) \\
((22, 56, 7 )) \\
((24, 24, 8 )) \\
((39, 24, 13))
\end{array}
$$

It would be very interesting to have an elegant combinatorial construction for
any of these codes.

(xiv) In Theorem~\ref{SB2} we listed all sets of
parameters of the form $[[n,n-2d+2,d]]$ for which an additive
code exists, and remarked that in each case the code is unique.
In \cite{RainsQd2} this result is extended to nonadditive codes.
In particular, any
$$((2,1,2)), ((4,4,2)) , ((5,2,3)), ((6,1,4))$$
code is equivalent to the unique
$$[[2,0,2]] , [[4,2,2]], [[5,1,3]], [[6,0,4]] ~~~~~$$
additive code, respectively.
On the other hand, for all $n > 2$, there exists a nonadditive
$((2n , 2^{2n-2} , 2))$ code.

(xv) There is a remarkable story behind this paper.
About a year and a half ago one of us (P.W.S.) was studying fault-tolerant
quantum computation,
and was led to investigate a certain group of $8 \times 8$ orthogonal
matrices.
P.W.S. asked another of us (N.J.A.S.) for the best method of
computing the order of this group.
N.J.A.S. replied by citing the computer algebra system MAGMA
\cite{Mag1}, \cite{Mag2}, \cite{Mag3}, and gave as an illustration the MAGMA
commands needed to specify a certain matrix group that had recently
arisen in connection with packings in Grassmannian spaces.
This group was the symmetry group of a packing of 70 4-dimensional subspaces of $\RR^8$ that had been discovered by computer search \cite{grass1}.
It too was an 8-dimensional group, of order 5160960.
To our astonishment the two groups turned out to be identical (not just isomorphic)!
We then discovered that this group was a member of an infinite
family of groups that played a central role in a joint paper \cite{CCKS96} of
another of the authors (A.R.C.).
This is the family of real Clifford groups $L_R$, described in Section~2
(for $n=3$, $L_R$ has order 5160960).

This coincidence led us to make connections which further advanced both
areas of research (fault-tolerant quantum computing \cite{ShoF} and
Grassmannian packings \cite{grass2}).

While these three authors were pursuing these investigations, the fourth
author (E.M.R.) happened to be present for a job interview and was able to
make further contributions to the Grassmannian packing problem
\cite{grass3}.  As the latter involved packings of $2^k$-dimensional
subspaces in $2^n$-dimensional space, it was natural to ask if the same
techniques could be used for constructing quantum-error-correcting codes,
which are also subspaces of $2^n$-dimensional space.  This question led
directly to \cite{CRSS96} and the present paper.  (Incidentally, he got the
job.)

A final postscript: At the 1997 IEEE International Symposium on Information Theory,
V.~I. Sidelnikov presented a paper ``On a finite group
of matrices generating orbit codes on the Euclidean sphere''
\cite{Sid2} (based on \cite{Sid1}, \cite{Kaz}).
It was no surprise to discover that --- although Sidelnikov did not identify them in this way ---
these were the Clifford groups appearing in yet another guise.
\subsection*{Acknowledgements}
\hsp
We thank Aaron Gulliver for finding the quasi-cyclic codes mentioned in Section~5, and our colleague Ronald H. Hardin for
his contributions to our search for interesting nonadditive codes.

\end{document}